\documentclass{emulateapj}
\usepackage[usenames]{color}

\begin{document}

\title{Star-forming or starbursting? The Ultraviolet Conundrum}


\author{M. Boquien\altaffilmark{1}, D. Calzetti\altaffilmark{1}, R. Kennicutt\altaffilmark{2,3}, D. Dale\altaffilmark{4}, C. Engelbracht\altaffilmark{3}, K. D. Gordon\altaffilmark{5}, S. Hong\altaffilmark{1}, J. C. Lee\altaffilmark{6,7} and J. Portouw\altaffilmark{3}}
\email{boquien@astro.umass.edu}

\altaffiltext{1}{University of Massachusetts, Department of Astronomy, LGRT-B 619E, Amherst, MA 01003, USA}
\altaffiltext{2}{Institute of Astronomy, University of Cambridge, Madingley Road, Cambridge CB3 0HA, UK}
\altaffiltext{3}{Steward Observatory, University of Arizona, Tucson, AZ 85721, USA}
\altaffiltext{4}{Department of Physics and Astronomy, University of Wyoming, Laramie, WY 82071, USA}
\altaffiltext{5}{STScI, 3700 San Martin Drive, Baltimore, MD 21218, USA}
\altaffiltext{6}{Carnegie Observatories, 813 Santa Barbara Street, Pasadena, CA 91101, USA}
\altaffiltext{7}{Hubble Fellow}

\begin{abstract}
Compared to starburst galaxies, normal star forming galaxies have been shown to display a much larger dispersion of the dust attenuation at fixed reddening through studies of the IRX-$\beta$ diagram (the IR/UV ratio ``IRX'' versus the UV color ``$\beta$''). To investigate the causes of this larger dispersion and attempt to isolate second parameters, we have used GALEX UV, ground-based optical, and Spitzer infrared imaging of 8 nearby galaxies, and examined the properties of individual UV and 24~$\mu$m selected star forming regions. We concentrated on star-forming regions, in order to isolate simpler star formation histories than those that characterize whole galaxies. We find that 1) the dispersion is not correlated with the mean age of the stellar populations, 2) a range of dust geometries and dust extinction curves are the most likely causes for the observed dispersion in the IRX-beta diagram 3) together with some potential dilution of the most recent star-forming population by older unrelated bursts, at least in the case of star-forming regions within galaxies, 4) we also recover some general characteristics of the regions, including a tight positive correlation between the amount of dust attenuation and the metal content. Although generalizing our results to whole galaxies may not be immediate, the possibility of a range of dust extinction laws and geometries should be accounted for in the latter systems as well.
\end{abstract}


\section{Introduction}
\label{sec:introduction}
Star formation is a fundamental process that governs the evolution of the baryonic matter in galaxies. It can deeply affect the host galaxy, its properties and appearance. The newly formed populations change the galaxy's spectral energy distribution (SED); the metals formed in the most massive stars pollute the interstellar medium and, in the presence of mechanical feedback from episodes of intense star formation, can be ejected from the galaxy and pollute the intergalactic medium. It is important to accurately measure star formation both as a function of spatial position within a galaxy and as a function of a galaxy's temporal evolution. The cosmic star formation rate density as a function of redshift \citep{madau1996a,hopkins2006b} has become a classic comparison benchmark, and a challenge, for both semi-analytic and hydrodynamical models of structure formation and evolution \citep[e.g.][]{katz1996a,hopkins2006a,stinson2006a,cattaneo2007a}. The youngest stellar populations, whose bolometric energy output is dominated by short-lived massive stars, emit the bulk of their energy in the restframe ultraviolet (UV). Since the pioneer work of \cite{donas1984a} the UV has become one of the main star formation rate (SFR) tracers used in modern Astronomy \citep{kennicutt1998a}. This regime, which has been previously difficult to observe in the local Universe, becomes the wavelength region ``par excellence'' to investigate star formation in galaxies at intermediate and high redshift. At these distances, the restframe UV is shifted to optical and near-infrared wavelengths, where current detectors afford the highest performance in terms of both sensitivity and angular resolution. Thus, it is no surprise that so far galaxy investigations in the redshift range z$\sim$2-7 have been dominated by restframe UV observations \citep{giavalisco2004a,bouwens2005a,mobasher2005a,sawicki2006a}. UV observations are the main canvass from which our current understanding of cosmic star formation is built.

For over a decade now, studies of galaxies at cosmological distances have employed the ``starburst attenuation curve'' \citep{calzetti1994a,calzetti2000a,meurer1999a} -- see next paragraph for a description of the physical conditions leading to such a law and equation 4 from \cite{calzetti2000a} -- to correct their restframe UV measurements for the effects of dust extinction \citep[e.g.][to name a few]{steidel1996a,steidel1999a,giavalisco2004a,daddi2005a,daddi2007a}. This curve is a powerful tool, in that it only requires information on the UV colors (or UV spectral slope) of a galaxy to provide an extinction correction, and recover the intrinsic UV flux. Assuming this curve, the typical UV extinction correction at redshift 3 is a factor $\sim5$ in luminosity \citep{steidel1999a}. However, in recent years, evidence has been accumulating that the starburst attenuation curve is not applicable to all galaxies \citep{buat2002a,buat2005a,noll2009a} and that the UV extinction is dependent on the age of the stellar populations \citep{cortese2008a}. More quiescently star-forming galaxies show characteristics in their UV SED that do not lend themselves to the same extinction ``treatment'' as more active starbursts. Applying the starburst attenuation curve to these more quiescent galaxies leads to up to an order of magnitude overestimate of the UV luminosity (and of the SFR). This result clearly represents a problem as deeper observations probe past the most powerful starburst to more normal galaxies with relatively lower SFR at increasingly higher redshifts. Along these lines, current determinations of UV luminosity functions at cosmological distances may be viewed with some suspicion, as all of them need some form of extinction correction \citep{steidel1999a,giavalisco2004a,sawicki2006a,teplitz2006a}.

One of the prescriptions of the starburst attenuation curve is based on a very simple, fully-empirical result. The ratio L(IR)/L(UV) is strongly correlated with the measured UV spectral slope $\beta$ (or the UV color) in local, UV-bright starburst galaxies \citep{meurer1999a,calzetti2000a}. The L(IR)/L(UV) ratio is a measure of the total dust opacity that affects the UV emission in a galaxy, or more generally, in a stellar system; the UV light absorbed by dust is re-emitted in the IR, independently of the details of the star-dust distribution, the dust properties, the metallicity, the stellar initial mass function, or the star formation history of the system \citep{gordon2000a,buat2005a}. Meurer et al.'s result, commonly called the IRX-$\beta$ relation, together with a host of previous results that showed strong correlations between $\beta$ and a number of dust reddening measures, form the basis for an extinction correction prescription that obviously applies to local (and possibly distant) starburst galaxies \citep{calzetti1994a,calzetti1996a,calzetti1997a,calzetti2001a}. The properties of the starburst attenuation curve imply specific conditions for the distribution of dust and stars within the starburst region \citep{calzetti1994a,calzetti1996a,witt2000a,charlot2000a,calzetti2001a}. The geometry that can account for all observed properties at UV, optical, near-IR, and far-IR wavelengths is that of a dust distribution foreground to the stellar population responsible for the bulk of the UV emission. The term foreground is here loosely used to indicate any geometry that puts the stars in the center of a shell-like dust distribution, and does not make any inference on the dust distribution (which can be homogeneous or clumpy). Starbursts provide a mechanism for creating such specific dust-star geometry: massive star wind and supernova feedback can create expanding bubbles of hot gas that sweep away a significant fraction of the dust and gas that resides in the starburst site itself. Within this scenario, the gas and dust end up at the ``edges'' of the starburst site thus creating an inhomogeneously distributed ``shell''. In addition, the starbust extinction curve also requires a specific type of dust grains lacking the 2175\AA\ bump \citep{gordon1997a} like that found in the SMC bar \citep{gordon1998a,gordon2003a}.

The starburst attenuation curve may thus not apply to a generic galaxy, as the coherent feedback mechanism present in starbursts may be missing in other galaxies. A number of studies have confirmed that more quiescent star-forming galaxies, although they generally trace a correlation in the IRX-$\beta$ plane, are also up to an order of magnitude less obscured than starbursts for similar $\beta$ values, and can have a larger dispersion, with a factor up to $\sim$5 larger spread around the mean correlation \citep{buat2002a,buat2005a,bell2002a,kong2004a,gordon2004a,seibert2005a,calzetti2005a,boissier2007a,dale2007a,munoz2009a}. UltraLuminous Infrared Galaxies (ULIRGs) also deviate from the starburst IRX-$\beta$ correlation \citep{goldader2002a}, and are located above it, i.e., opposite to the star-forming galaxies. ULIRGs are dustier, more compact, and denser systems than the UV-bright starbursts; therefore, their location in the IRX-$\beta$ diagram is plausibly accounted for by an ``homogeneous'' mixture of dust and stars \citep{calzetti2001a}. ULIRGs likely represent less of a concern for the use of the UV emission as a SFR tracer, as in general they will not be detectable at restframe UV wavelengths in cosmological samples. Quiescent star-forming galaxies are a far larger concern, as they are relatively bright in the UV and become increasingly detected as surveys go deeper and, by the nature of the luminosity function, are more numerous than starbursts. Alas, for these galaxies, the next step in the investigation, understanding and quantifying the nature of the deviation from the starbursts' IRX-$\beta$ correlation, has so far eluded any attempt. \cite{bell2002a} suggested that a combination of variations in the dust geometry and the characteristics (age, etc.) of the stellar populations could be the reason for the deviation observed in the HII regions of the Large Magellanic Cloud. The sparse data available, however, prevented further insights into the problem. Similar limitations were encountered by \cite{gordon2004a} in their analysis of the star-forming regions in M~81. \cite{kong2004a}, using a non-homogeneous UV dataset, suggested that the deviations of a large sample of quiescent star-forming galaxies from the starburst relation could be parametrized by the b-parameter, the ratio of the current to past average star formation rate. However, neither \cite{seibert2005a}, using the more homogeneous GALEX All-sky Imaging Survey sample plus IR data from IRAS, nor \cite{johnson2007b}, using SDSS data, Spitzer, and spectroscopic observations, confirmed Kong et al.'s suggestion. Some of these works have also produced ``mean'' obscuration curves for quiescent star-forming galaxies, but the intrinsic spread of the data around these mean curves is more than an order of magnitude, versus $\sim$2-3 typical of the starbursts. This suggests the existence of a ``second parameter'' for the star-forming galaxies \citep{kong2004a}, which needs to be identified and understood if the UV is going to be used as a tracer of SFR.

In this paper, we investigate the possible reason for the deviation from the starburst IRX-$\beta$ relation concentrating on sub-kpc star forming regions within galaxies, which provide simpler samples of stellar population than whole galaxies. We use GALEX observations to measure the UV luminosity, Spitzer ones to derive the total infrared luminosity and optical ones to estimate the age.

In section 2 we present the sample of galaxies selected for this study, in section 3 we present the UV, IR and optical observations, in section 4 we present the data processing and photometry methods carried out, in section 5 we present how we model the star forming regions, in section 6 we proceed on to test the validity of the age as the ``second parameter'', in section 7 we test different extinction laws, in section 8 we test the metallicity, we discuss the results presented in section 9 and we finally conclude in section 10.

\section{Sample of galaxies}
Since we wish to investigate individual star forming regions within galaxies, we select galaxies where such regions can be resolved. Indeed the typical resolution in mid-infrared at 24~$\mu$m with Spitzer MIPS or ultraviolet with GALEX -- which trace star formation -- is about 5-6\arcsec. The typical size of a giant HII complex, about 300~pc can be resolved at a distance up to 10~Mpc at this resolution. Within this size at most a few stellar clusters may reside and contribute to the ultraviolet and infrared emission. Thus this constraint on the distance ensures that there are as few clusters as possible in an individual giant HII region. On the other hand we also exclude very nearby galaxies that would provide a very high level of detail. We have applied the following criteria to select the sample of galaxies for this study: 1) a distance closer than 10~Mpc, 2) to be star-forming spiral galaxies, 3) to cover a wide range in total SFR and SFR surface density, 4) not to be too inclined (minor axis/major axis $> 0.5$), 5) have GALEX, Spitzer data available.

The final sample consists of 8 star-forming galaxies, details are given in Table \ref{tab:galaxy-parameters}. Their star formation rates range from 0.004~M$_\sun$~yr$^{-1}$~kpc$^{-2}$ to 0.046~M$_\sun$~yr$^{-1}$~kpc$^{-2}$ with an average of $0.020\pm0.021$~M$_\sun$~yr$^{-1}$~kpc$^{-2}$ calculated over the area defined by $R_{25}$.
{\tabletypesize{\tiny}
\begin{deluxetable*}{lccccccccccc}
\tablecolumns{12}
\tablewidth{0pc}
\tablecaption{Galaxy parameters\label{tab:galaxy-parameters}}
\tablehead{
\colhead{Name}&\colhead{Morphology}&\colhead{$\alpha$\tablenotemark{a}}&\colhead{$\delta$\tablenotemark{a}}&\colhead{Distance}&\colhead{PA\tablenotemark{b}}&\colhead{Major axis\tablenotemark{a}}&\colhead{Minor axis\tablenotemark{a}}&\colhead{$\rho_{25}$\tablenotemark{c,d}}&\colhead{[O/H]\tablenotemark{d}}&\colhead{$\Delta$[O/H]\tablenotemark{d}}&\colhead{[NII]/H$\alpha$}\\
\colhead{}&\colhead{}&\colhead{(J2000.0)}&\colhead{(J2000.0)}&\colhead{(Mpc)}&\colhead{($\deg$)}&\colhead{(\arcmin)}&\colhead{(\arcmin)}&\colhead{(\arcmin)}&\colhead{($\rho=0$)}&\colhead{(dex/$\rho_{25}$)}&}\\
\startdata
\textcolor{red}{NGC~0628} (M~74)&SAbc&01 36 41.8&+15 47 01&7.3&\phantom{00}0&10.5&\phantom{0}9.5&\phantom{0}5.24&9.16&$-0.58$&0.47\\
\textcolor{Brown}{NGC~2403}&SABcd&07 36 51.4&+65 36 09&3.13&123&21.9&12.3&10.94&8.88&$-0.27$&0.63\\
\textcolor{green}{NGC~3031} (M~81)&SAab&09 55 33.2&+69 03 55&3.55&155&26.9&14.1&13.46&9.08&$-0.43$&0.28\\
\textcolor{blue}{NGC~5055} (M~63)&SAbc&13 15 49.3&+42 01 45&7.8&107&12.6&\phantom{0}7.2&\phantom{0}6.30&9.27&$-0.54$&\nodata\\
\textcolor{magenta}{NGC~5194} (M~51)&SABbc&13 29 52.7&+47 11 43 &7.62&22&11.2&\phantom{0}6.9&\phantom{0}5.61&9.30&$-0.49$&0.50\\
\textcolor{Orange}{NGC~5236} (M~83)&SABc&13 37 00.9&$-$29 51 56&4.5&0&12.9&11.5&5.9&9.24&$-0.19$&0.50\\
\textcolor{black}{NGC~5457} (M~101)&SABcd&14 03 12.6&+54 20 57&7.2&\phantom{0}39&28.8&26.9&14.1&8.89&$-0.93$&\nodata\\
\textcolor{cyan}{NGC~7793}&SAd&23 57 49.8&$-$32 35 28&3.91&102&\phantom{0}9.3&\phantom{0}6.3&\phantom{0}4.67&8.93&$-0.32$&0.28\\
\enddata
\tablenotetext{a}{Data from the NASA/IPAC Extragalactic Database.}
\tablenotetext{b}{The position angle is calculated manually adjusting an ellipse to the galaxy in the R band image (if available).}
\tablenotetext{c}{Radius of the major axis at the $\mu_B$=25~mag~arcsec$^{-2}$ isophote.}
\tablenotetext{d}{Data obtained from \cite{zaritsky1994a} and Moustakas et al. (2009, in preparation).}
\end{deluxetable*}
}
\section{Observations}

\subsection{GALEX}

We have obtained far and near ultraviolet processed count and intensity maps from the GALEX archives. Thanks to its large field of view of 1.24$^\circ$, it can encompass the full extent of the selected galaxies with a single pointing. The far ultraviolet band is centered on $\lambda_\textrm{eff}$=151~nm whereas the near ultraviolet one is centered on $\lambda_\textrm{eff}$=227~nm. Detailed information about the instruments are provided in \cite{martin2005a}.

Galaxies have been observed as part of the Nearby Galaxy Survey (NGC~0628, NGC~2403, NGC~5055, NGC~5194, NGC~5457 and NGC~7793) and Guest investigators programs (program 71 cycle 1 for NGC~3031 and program 5 in cycle 3 for NGC~5236).

\subsection{Ground based optical}

Ground based optical images consist of narrow-band H$\alpha$ observations as well as broad band U, B and R observations.

Most H$\alpha$ images have been obtained from the Spitzer Infrared Nearby Galaxies Survey \citep[SINGS,][]{kennicutt2003a} archives, fifth delivery, final enhanced products released on 2007-04-10. Observations have been carried out at the National Optical Astronomy Observatory, in 2001 on the Kitt Peak National Observatory (KPNO) 2.1~m and Cerro Tololo Inter-American Observatory (CTIO) 1.5~m telescopes. More details are available in the SINGS ``User's Guide''\footnote{\url{http://data.spitzer.caltech.edu/popular/sings/\\20070410\_enhanced\_v1/Documents/sings\_fifth\_delivery\_v2.pdf}}. For \textcolor{Orange}{NGC~5236}, the data were obtained on the Bok 2.3~m telescope on 2005-05-10.

Continuum emission has been subtracted from the H$\alpha$ image using the corresponding scaled R band image following the method described in Hong et al. (in preparation).

SINGS spectroscopic observations (Moustakas et al., in preparation) have been used to infer the mean [NII]/[H$\alpha$] ratio within the galaxies. The measured fluxes have been corrected for contamination by the [NII] lines accordingly. As no spectrum was available for the high-metallicity galaxy \textcolor{Orange}{NGC~5236}, we have assumed [NII]/[H$\alpha$]=0.5, close to 0.53 found by \cite{kennicutt2008a} which has been estimated based on a scaling relationship between M(B) and [NII] as described in appendix B of that paper. Although the relative strength of the [NII] lines can change from one HII region to another and can evolve as a function of the radial distance, we have assumed a constant ratio for each galaxy. Indeed very super-metallic and sub-metallic star forming regions -- in which the relative strength of the [NII] lines may deviate significantly from the mean observed values -- are less likely to be detected in the UV for the former and in the mid-infrared for the latter.

Images in the U, B and R bands have been obtained either from the SINGS archives -- as mentioned in the previous section -- or observed in the context of the Steward Nearby U-band GaLaxy (SNUGL) survey, using 90prime \citep{williams2004a}.

\subsection{Spitzer}

Spitzer is a key observatory to estimate the total infrared luminosity. Indeed its two imagers cover a large range of infrared emission, from 3.6~$\mu$m to 8.0~$\mu$m for IRAC and 24~$\mu$m to 160~$\mu$m for MIPS. The relations to estimate the bolometric infrared luminosity routinely use the IRAC 8.0~$\mu$m and the MIPS 24~$\mu$m bands \citep[see for instance][ Boquien et al. in preparation, and references therein]{calzetti2005a,perez2006a}. All galaxies but two have been observed as part of SINGS. We use the processed legacy data provided by the SINGS team. The only galaxies that are not observed by SINGS are \textcolor{Orange}{NGC~5236} \cite[also observed by][]{engelbracht2008a} and \textcolor{black}{NGC~5457} \citep[observed by][]{gordon2008a} are part of the Spitzer Local Volume Legacy Survey (Dale et al., 2009).

\subsection{Summary}

\begin{figure}[!ht]
\includegraphics[width=\columnwidth]{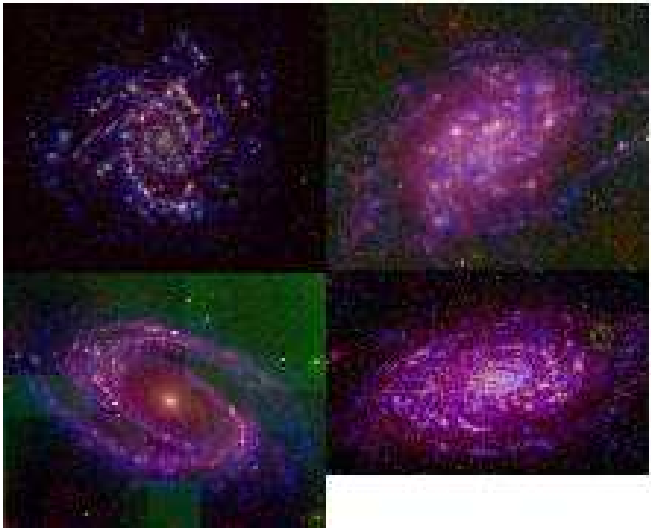}
\caption{Combined FUV (blue channel), H$\alpha$ (green channel) and 24~$\mu$m (red channel). The top-left image in NGC~0628, the top-right one NGC~2403, the bottom-left one NGC~3031 and the bottom-right one NGC~5055. Note that the H$\alpha$ image of NGC~3031 does not cover all the galaxy and is the combination of several pointings which have slightly different background levels can can be clearly seen in the image.\label{fig:tumbnails-1}}
\end{figure}

\begin{figure}[!ht]
\includegraphics[width=\columnwidth]{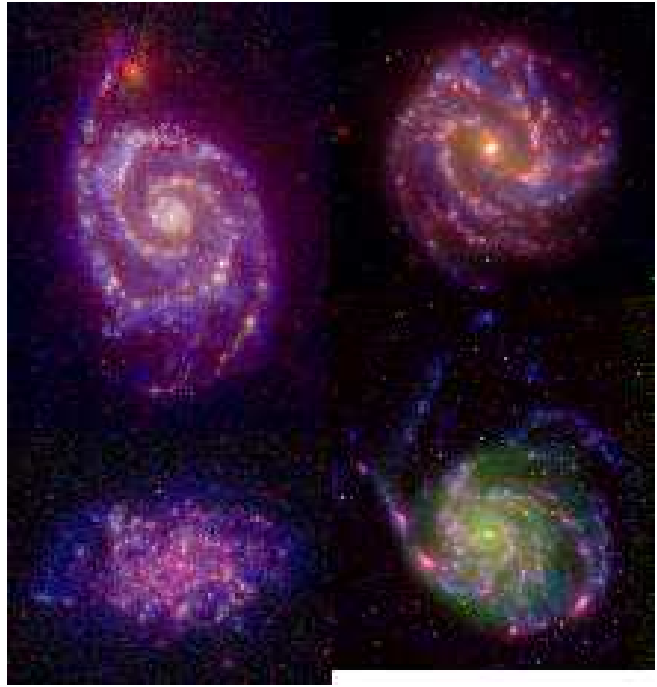}
\caption{Combined FUV (blue channel), H$\alpha$ (green channel) and 24~$\mu$m (red channel). The top-left image in NGC~5194, the top-right one NGC~5236, the bottom-left one NGC~7793 and the bottom-right one NGC~5457.\label{fig:tumbnails-2}}
\end{figure}

Images of the galaxies in FUV, H$\alpha$ and 24~$\mu$m are presented in Figures \ref{fig:tumbnails-1} and \ref{fig:tumbnails-2}. For \textcolor{Orange}{NGC~5236} the missing H$\alpha$ image has been substituted with the R band one.

A summary of the bands, telescopes used, observing dates, exposure times and sensitivities of all observations are presented in Table \ref{tab:obs-summary}.

\begin{deluxetable*}{cccccc}
\tablecolumns{6}
\tablewidth{0pc}
\tablecaption{Images\label{tab:obs-summary}}
\tablehead{
\colhead{Galaxy}&\colhead{Band}&\colhead{Telescope}&\colhead{Date}&\colhead{Exposure time}&\colhead{1-$\sigma$~sensitivity}\\
\colhead{}&\colhead{}&\colhead{}&\colhead{}&\colhead{(s)}&\colhead{(nJy arcsec$^{-2}$)\tablenotemark{a}}
}
\startdata
\textcolor{red}{NGC~0628}&FUV&GALEX&2003-10-07&1636&31\\
\textcolor{red}{\nodata}&NUV&\nodata&\nodata&\nodata&28\\
\textcolor{red}{\nodata}&U&Bok 2.3~m&2004-12-12&600&42\\
\textcolor{red}{\nodata}&B&CTIO 1.5m&2001-10-20&480&50\\
\textcolor{red}{\nodata}&R&\nodata&2001-10-20&180&189\\
\textcolor{red}{\nodata}&H$\alpha$&\nodata&2001-10-21&600&$1.36\times10^{-20}$\\
\textcolor{red}{\nodata}&8~$\mu$m&Spitzer/IRAC&2004-07-28&240&1245\\
\textcolor{red}{\nodata}&24~$\mu$m&Spitzer/MIPS&2005-01-23&160&2515\\
\textcolor{Brown}{NGC~2403}&FUV&GALEX&2003-12-05&3317&22\\
\textcolor{Brown}{\nodata}&NUV&\nodata&\nodata&\nodata&18\\
\textcolor{Brown}{\nodata}&U&Bok 2.3~m&2004-12-09&600&35\\
\textcolor{Brown}{\nodata}&B&\nodata&2004-12-10&300&47\\
\textcolor{Brown}{\nodata}&R&KPNO 2.1m&2001-11-16&180&74\\
\textcolor{Brown}{\nodata}&H$\alpha$&\nodata&2001-11-15&600&$6.70\times10^{-21}$\\
\textcolor{Brown}{\nodata}&8~$\mu$m&Spitzer/IRAC&2004-10-12&240&790\\
\textcolor{Brown}{\nodata}&24~$\mu$m&Spitzer/MIPS&2004-10-12&160&1404\\
\textcolor{green}{NGC~3031}&FUV&GALEX&2006-01-05&11912&23\\
\textcolor{green}{\nodata}&NUV&\nodata&2005-01-12&\nodata&17\\
\textcolor{green}{\nodata}&U&Bok 2.3~m&2004-12-12&600&42\\
\textcolor{green}{\nodata}&B&\nodata&2004-05-10&300&48\\
\textcolor{green}{\nodata}&R&KPNO 2.1m&2001-03-30&420&146\\
\textcolor{green}{\nodata}&H$\alpha$&\nodata&2001-03-30&900&$1.33\times10^{-20}$\\
\textcolor{green}{\nodata}&8~$\mu$m&Spitzer/IRAC&2004-05-01&240&1219\\
\textcolor{green}{\nodata}&24~$\mu$m&Spitzer/MIPS&2004-10-16&160&1597\\
\textcolor{blue}{NGC~5055}&FUV&GALEX&2004-03-31&1655&23\\
\textcolor{blue}{\nodata}&NUV&\nodata&\nodata&\nodata&23\\
\textcolor{blue}{\nodata}&U&Bok 2.3~m&2005-02-09&600&47\\
\textcolor{blue}{\nodata}&B&\nodata&\nodata&300&60\\
\textcolor{blue}{\nodata}&H$\alpha$&KPNO 0.9m&1999-04-20&420&\\
\textcolor{blue}{\nodata}&8~$\mu$m&Spitzer/IRAC&2004-12-15&240&1399\\
\textcolor{blue}{\nodata}&24~$\mu$m&Spitzer/MIPS&2005-01-26&160&1750\\
\textcolor{magenta}{NGC~5194}&FUV&GALEX&2003-06-19&2520&27\\
\textcolor{magenta}{\nodata}&NUV&\nodata&\nodata&\nodata&20\\
\textcolor{magenta}{\nodata}&U&Bok 2.3~m&2004-06-20&600&20\\
\textcolor{magenta}{\nodata}&B&KPNO 2.1m&2001-03-28&720&37\\
\textcolor{magenta}{\nodata}&R&\nodata&2001-03-28&360&61\\
\textcolor{magenta}{\nodata}&H$\alpha$&\nodata&2001-03-28&900&$5.04\times10^{-21}$\\
\textcolor{magenta}{\nodata}&8~$\mu$m&Spitzer/IRAC&2004-05-22&240&1007\\
\textcolor{magenta}{\nodata}&24~$\mu$m&Spitzer/MIPS&2004-06-22&160&1701\\
\textcolor{Orange}{NGC~5236}&FUV&GALEX&2003-06-07&1343&50\\
\textcolor{Orange}{\nodata}&NUV&\nodata&\nodata&\nodata&35\\
\textcolor{Orange}{\nodata}&U&Bok~2.3m&2005-05-09&600&88\\
\textcolor{Orange}{\nodata}&B&\nodata&\nodata&300&104\\
\textcolor{Orange}{\nodata}&R&\nodata&2005-05-10&300&125\\
\textcolor{Orange}{\nodata}&H$\alpha$&\nodata&2005-05-10&600&$1.37\times10^{-20}$\\
\textcolor{Orange}{\nodata}&8~$\mu$m&Spitzer/IRAC&2008-03-05&240&1496\\
\textcolor{Orange}{\nodata}&24~$\mu$m&Spitzer/MIPS&2006-02-18&160&2005\\
\textcolor{black}{NGC~5457}&FUV&GALEX&2003-06-20&1040&38\\
\textcolor{black}{\nodata}&NUV&\nodata&\nodata&\nodata&30\\
\textcolor{black}{\nodata}&U&Bok 2.3~m&2005-05-09&600&49\\
\textcolor{black}{\nodata}&B&\nodata&2005-05-10&300&31\\
\textcolor{black}{\nodata}&8~$\mu$m&Spitzer/IRAC&2004-03-08&85&1212\\
\textcolor{black}{\nodata}&24~$\mu$m&Spitzer/MIPS&2004-05-10&200&1398\\
\textcolor{cyan}{NGC~7793}&FUV&GALEX&2003-09-13&1553&26\\
\textcolor{cyan}{\nodata}&NUV&\nodata&\nodata&\nodata&26\\
\textcolor{cyan}{\nodata}&U&Danish 1.54m&1997-06-09&900&63\\
\textcolor{cyan}{\nodata}&B&CTIO 1.5m&2001-10-18&480&73\\
\textcolor{cyan}{\nodata}&R&\nodata&2001-10-18&180&143\\
\textcolor{cyan}{\nodata}&H$\alpha$&\nodata&2001-10-18&600&$1.43\times10^{-20}$\\
\textcolor{cyan}{\nodata}&8~$\mu$m&Spitzer/IRAC&2004-06-10&240&1115\\
\textcolor{cyan}{\nodata}&24~$\mu$m&Spitzer/MIPS&2004-07-10&160&1829\\
\enddata
\tablenotetext{a}{In units of W~m$^{-2}$~s$^{-1}$~arcsec$^{-2}$ for H$\alpha$ images.}
\end{deluxetable*}

\section{Data processing and photometry}

\subsection{Definition of the apertures\label{ssec:apertures}}

Given the multi-wavelength nature of this study, we have selected the star forming regions using the GALEX FUV and the Spitzer MIPS 24~$\mu$m images. Indeed, not only do those wavelengths trace star formation but they also have a similar resolution ($\sim$5-6\arcsec) which ensures that the totality of the region is encompassed in other wavelengths that provide a better resolution such as the IRAC 8.0~$\mu$m and the optical bands. Given that they offer a complementary view of star formation -- far ultraviolet emission absorbed by the dust is reemitted in the infrared --, the selection of the star forming regions has been performed independently in those bands. This allows us not to miss regions that have a very low or a very high extinction, the former being mainly visible in FUV while the latter are primarily detected in the infrared.

Even at a resolution of $\sim$5-6\arcsec, star forming regions complexes are sufficiently resolved that their shape have to be taken into account using polygonal apertures. Visual inspection shows that the irregular shapes are mainly due to the overlap of very close star forming regions. The apertures have been made as small as possible to reduce the number of star forming regions inside an individual aperture to limit the contamination by nearby sources. The goal is to select star forming regions as simple as possible, and to avoid the most complex ones composed of stellar populations of different ages. To achieve this goal, we have defined apertures using polygons. Each polygon is defined manually to encompass as few star forming regions as possible within a single aperture as well as to avoid artifacts, background and foreground objects.

Even if some regions are most likely selected in the UV and in the IR but with different apertures, no attempt has been made to make a link between the two. This is made difficult by the fact that for a given region the apertures can be quite different, cover only a part of the region or be split into several apertures. However, a visual inspection of the apertures shows that some regions have an aperture defined only from either UV or IR even if it is visible in both bands. This is due to the fact that the apertures for each band are built mainly for the most prominent regions.

The background flux is defined on a per-aperture basis using an annulus around the aperture. The radius and the width of this annulus is defined for each galaxy depending on the surface density of star forming regions. Local background subtraction has been performed in all bands. The background level is then calculated using the mode of the pixels enclosed in the background annulus. To improve the determination of the galactic background level, for some wavelengths we have invalidated emission above a given threshold in the background annulus. This threshold has been determined manually to eliminate as much emission as possible from the nearby sources while keeping the genuine background emission. This was done for the GALEX bands as well as the Spitzer IRAC 8.0~$\mu$m and MIPS 24~$\mu$m bands. This threshold was not applied to broadband optical images where young star forming regions are much less prominent ; in the U, B and R bands the local background is dominated by an evolved stellar population.

The raw fluxes and the errors have been calculated using the IRAF procedure {\sc polyphot}.

Following the aforementioned criteria, we have selected a total of 1146 star forming regions in the 8 galaxies constituting the sample. In order to increase the significance of the results we have subsequently selected only star forming regions which have flux uncertainties of at most 20\% in the GALEX FUV, NUV and the Spitzer MIPS 24~$\mu$m band, leaving us with a total sample of 324 regions. The number of UV selected and infrared selected regions is presented in Table \ref{tab:number-regions}, for each galaxy in our sample.

\begin{deluxetable*}{ccccccccc}
\tablecolumns{9}
\tablewidth{0pc}
\tablecaption{Number of selected star forming regions\label{tab:number-regions}}
\tablehead{
\colhead{Galaxy}&\multicolumn{2}{c}{UV selected regions}&\colhead{}&\multicolumn{2}{c}{IR selected regions}&\colhead{}&\multicolumn{2}{c}{Regions}\\
\cline{2-3} \cline{5-6} \cline{8-9} \\
\colhead{}&\colhead{Total}&\colhead{Selected}&\colhead{}&\colhead{Total}&\colhead{Selected}&\colhead{}&\colhead{Total}&\colhead{Selected}}
\startdata
\textcolor{red}{NGC~0628}&\phantom{0}57&26&&\phantom{0}36&\phantom{0}8&&\phantom{0}93&34\\
\textcolor{Brown}{NGC~2403}&\phantom{0}80&28&&\phantom{0}24&12&&104&40\\
\textcolor{green}{NGC~3031}&135&23&&\phantom{0}64&13&&199&36\\
\textcolor{blue}{NGC~5055}&\phantom{0}51&\phantom{0}7&&\phantom{0}40&\phantom{0}5&&\phantom{0}91&12\\
\textcolor{magenta}{NGC~5194}&\phantom{0}74&32&&\phantom{0}63&25&&137&57\\
\textcolor{Orange}{NGC~5236}&\phantom{0}67&30&&\phantom{0}52&22&&119&52\\
\textcolor{black}{NGC~5457}&143&50&&124&34&&267&84\\
\textcolor{cyan}{NGC~7793}&\phantom{0}66&6&&\phantom{0}70&\phantom{0}3&&136&\phantom{0}9\\
\hline
Total&673&202&&473&122&&1146&{\bf 324}
\enddata
\end{deluxetable*}

\subsection{Aperture correction}

We have performed an aperture correction on the infrared fluxes. As the method routinely used is not tailored for non circular apertures, we have proceeded as described in \citet{boquien2007a}. We have calculated an ``equivalent radius'': $r=\sqrt{S/\pi}$, $r$ being the equivalent radius and $S$ the surface of the aperture in units of pixels. The flux has then been corrected using this equivalent radius combined with the formulas provided in the Spitzer handbook for extended sources for IRAC\footnote{Also available at the following address: \url{http://ssc.spitzer.caltech.edu/irac/calib/extcal/}.} and in \cite{engelbracht2007a} for MIPS. For the 8.0~$\mu$m (24$\mu$m) band the average correction factor was $0.94\pm0.17$ (resp. $1.70\pm 2.11$). No aperture correction has been performed neither on UV images -- as the corrections are very small -- nor on optical images, thanks to their higher angular resolution. 

\subsection{Correction of the Galactic extinction}

We have corrected our UV and optical data for the effect of foreground Galactic extinction using the \citet{cardelli1989a} attenuation curve along with the differential extinction $E\left(B-V\right)$ provided by the NASA/IPAC extragalactic database.

\subsection{Fluxes and main properties of the sample}

\tabletypesize{\tiny}
\begin{deluxetable*}{cccccccccccc}
\tablecolumns{12}
\tablewidth{0pc}             
\tablecaption{Apertures positions and fluxes\label{tab:fluxes}}
\tablehead{
\colhead{ID}&\colhead{$\alpha$\tablenotemark{b}}&\colhead{$\delta$\tablenotemark{b}}&\colhead{FUV\tablenotemark{c}}&\colhead{NUV\tablenotemark{c}}&\colhead{U\tablenotemark{c}}&\colhead{B\tablenotemark{c}}&\colhead{R\tablenotemark{c}}&\colhead{8~$\mu$m\tablenotemark{de}}&\colhead{24~$\mu$m\tablenotemark{d}}&\colhead{H$\alpha$\tablenotemark{c}}&\colhead{R$_e$}\\
&\colhead{(J2000.0)}&\colhead{(J2000.0)}&\colhead{(mJy)}&\colhead{(mJy)}&\colhead{(mJy)}&\colhead{(mJy)}&\colhead{(mJy)}&\colhead{(mJy)}&\colhead{(mJy)}&\colhead{$10^{-17}$~W~m$^{-2}$~s$^{-1}$}&\colhead{\arcsec}}
\startdata
NGC~0628-U-01&01 36 28.0&+15 46 59&  0.14$\pm$  0.02&  0.13$\pm$  0.03&  0.20$\pm$  0.05&  0.23$\pm$  0.07&  0.17$\pm$  0.13&  2.81$\pm$  0.52&  4.52$\pm$  0.66&  3.74$\pm$  0.74& 15.3\\                                                                                                                                                                      
NGC~0628-U-02&01 36 29.1&+15 48 21&  0.19$\pm$  0.01&  0.21$\pm$  0.02&  0.38$\pm$  0.04&  0.34$\pm$  0.06&  0.37$\pm$  0.12& 11.96$\pm$  0.40& 32.47$\pm$  0.61& 11.86$\pm$  0.62& 10.0\\                                                                                                                                                                      
NGC~0628-U-03&01 36 30.3&+15 49 22&  0.29$\pm$  0.05&  0.35$\pm$  0.05&  0.86$\pm$  0.10&  1.64$\pm$  0.14&  1.94$\pm$  0.30&  7.24$\pm$  1.01&  8.67$\pm$  1.48&  5.49$\pm$  1.68& 11.9\\                                                                                                                                                                      
NGC~0628-U-04&01 36 36.1&+15 48 06&  0.17$\pm$  0.03&  0.22$\pm$  0.03&  0.31$\pm$  0.06&  0.41$\pm$  0.11&  0.36$\pm$  0.15&  6.30$\pm$  0.89&  9.81$\pm$  1.12&  3.06$\pm$  0.74&  7.3\\                                                                                                                                                                      
NGC~0628-U-05&01 36 36.2&+15 47 26&  0.42$\pm$  0.07&  0.48$\pm$  0.08&  0.69$\pm$  0.18&  0.97$\pm$  0.39&  0.90$\pm$  0.60& 21.41$\pm$  2.32& 18.28$\pm$  2.04&  5.95$\pm$  1.67& 12.3\\
NGC~0628-U-06&01 36 36.4&+15 50 17&  0.55$\pm$  0.10&  0.57$\pm$  0.10&  0.95$\pm$  0.18&  1.03$\pm$  0.31&  0.96$\pm$  0.53& 13.90$\pm$  2.61& 24.61$\pm$  4.09& 22.13$\pm$  2.90& 12.8\\
NGC~0628-U-07&01 36 37.1&+15 48 09&  0.35$\pm$  0.05&  0.34$\pm$  0.05&  0.55$\pm$  0.12&  0.70$\pm$  0.24&  0.74$\pm$  0.29& 26.09$\pm$  3.14& 45.01$\pm$  2.90& 13.02$\pm$  1.38&  8.1\\
NGC~0628-U-08&01 36 37.8&+15 45 01&  0.55$\pm$  0.06&  0.66$\pm$  0.07&  0.85$\pm$  0.12&  0.95$\pm$  0.22&  0.82$\pm$  0.30& 21.96$\pm$  1.70& 70.66$\pm$  2.14& 17.26$\pm$  1.56& 16.2\\
NGC~0628-U-09&01 36 38.8&+15 44 12&  1.97$\pm$  0.18&  2.14$\pm$  0.19&  2.60$\pm$  0.30&  3.02$\pm$  0.58&  2.34$\pm$  0.81& 26.32$\pm$  2.94& 76.63$\pm$  2.64& 37.59$\pm$  4.51&  8.5\\
NGC~0628-U-10&01 36 38.8&+15 49 13&  0.69$\pm$  0.04&  0.83$\pm$  0.04&  1.05$\pm$  0.08&  1.50$\pm$  0.13&  1.16$\pm$  0.29&  7.95$\pm$  0.82& 13.19$\pm$  0.83&  6.45$\pm$  1.18& 17.1\\
\enddata
\tablecomments{Units of right ascension are hours, minutes, and seconds, and units of declination are degrees, arcminutes, and arcseconds. Aperture correction has only been applied to infrared bands. Galactic extinction correction has been performeed on UV and optical fluxes. Table \ref{tab:fluxes} is published in its entirety in the electronic edition of the Astrophysical Journal. A portion is shown here for guidance regarding its form and content.}
\tablenotetext{a}{Format: names of the galaxy-band of selection-number. U (respectively I) indicates that the aperture has been defined from the FUV (resp. 24~$\mu$m) image.}
\tablenotetext{b}{Position of the barycenter of the apertures.}
\tablenotetext{c}{Stellar and ionized gas fluxes have been corrected for the Galactic foreground extinction using the \cite{cardelli1989a} law.}
\tablenotetext{d}{Infrared fluxes have been corrected for aperture effects.}
\tablenotetext{e}{Stellar emission in the 8~$\mu$m has been subtracted using the relation presented in \cite{helou2004a}: $F_\nu(8\ PAH)=F_\nu\left(8\right)-0.232\times F_\nu\left(3.6\right)$.}
\end{deluxetable*}

\begin{figure}[!ht]
\includegraphics[width=\columnwidth]{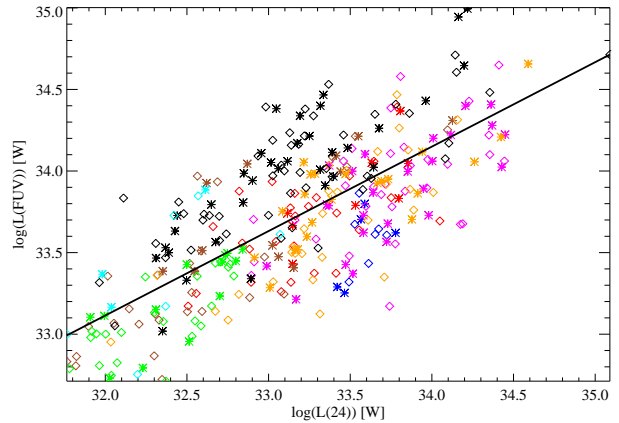}
\caption{Far ultraviolet luminosity versus the 24~$\mu$m one. The regions selected in ultraviolet (resp. mid-infrared) are represented by diamonds (resp. by asterisks). The color indicates in which galaxy the star forming regions are located: \textcolor{red}{NGC~0628: red} ; \textcolor{Brown}{NGC~2403: brown} ; \textcolor{green}{NGC~3031: green} ; \textcolor{blue}{NGC~5055: blue} ; \textcolor{magenta}{NGC~5194: magenta} ; \textcolor{Orange}{NGC~5236: Orange} ; \textcolor{black}{NGC~5457: black} ; \textcolor{cyan}{NGC~7793: cyan}. The black solid line represents the fit for the whole sample.\label{fig:24_vs_fuv}}
\end{figure}

The observed fluxes in all bands are presented in Table \ref{tab:fluxes}. The luminosities of the star forming regions span a factor $\sim$200 in FUV ($5.15\times10^{32}$ to $1.00\times10^{35}$~W) and a factor $\sim$2000 in 24~$\mu$m ($5.81\times10^{31}$ to $1.23\times10^{35}$~W). We can see in Figure~\ref{fig:24_vs_fuv} that those two star formation tracers are strongly correlated with a Pearson correlation coefficient $r=0.75$ between $\log\left(L\left(24\right)\right)$ and $\log\left(L\left(FUV\right)\right)$.

We observe that especially at higher luminosities, the 24~$\mu$m luminosity increases faster than the FUV one. This is most likely an extinction effect, as can be seen in infrared luminous galaxies for instance. The fit of the sample gives $\log L\left(FUV\right)=16.55\pm0.85+\left(0.52\pm0.03\right)\times \log L(24)$. The slope is as expected significantly smaller than 1.

\begin{figure}[!ht]
\includegraphics[width=\columnwidth]{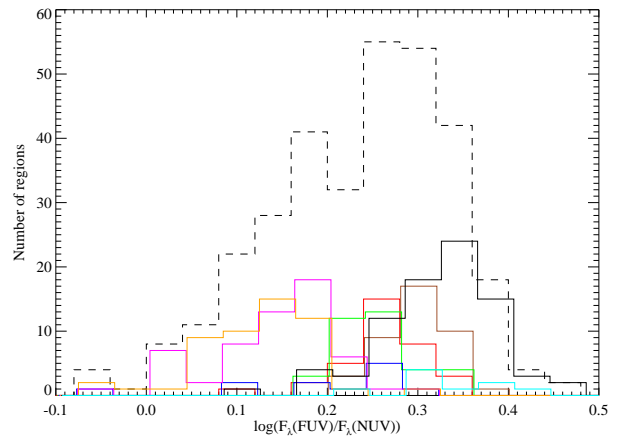}
\caption{Distribution of the UV colors as defined by $\log F_\lambda\left(FUV\right)/F_\lambda\left(NUV\right)$. The dashed line histogram represents the whole sample. The colored lines represent the histogram for each galaxy. The bin width is 0.04 dex. Histograms are slightly offset horizontally from each other by a small fraction of their bin width for better readability. The colors are the same as defined in Figure~\ref{fig:24_vs_fuv}.\label{fig:UV-colors}}
\end{figure}

While the correlation between the FUV and the 24~$\mu$m star formation tracers is good, the UV bands are impacted by the dust. The distribution of UV colors is presented in Figure~\ref{fig:UV-colors}.

There is a clear distinction in the UV colors distribution among the galaxies. The bluest one \textcolor{black}{NGC~5457} peaks 0.2 dex bluer than the reddest one, \textcolor{Orange}{NGC~5236}. This may be a consequence of the presence of older UV-emitting stars unrelated to the current star formation episode or the extinction which affects UV luminosities (a range of 0.2 dex corresponds to an extinction range $A_V\simeq1$ for the stellar component assuming a starburst attenuation curve). This is supported by the fact that very few star forming regions have an UV color that is compatible with an ionizing population even though H$\alpha$ is detected in most of them which indicates a reddening unrelated to the age of the current episode. Indeed, assuming a \cite{kroupa2001a} IMF (Initial Mass Function), and a solar metallicity, a simple stellar population will still have EW(H$\alpha$)$\simeq$1\AA\ after $20\times10^6$ years.

Among the 324 selected star forming regions, we have suitable H$\alpha$ imaging information for 228 and we measure their fluxes. The lack of H$\alpha$ observations for \textcolor{blue}{NGC~5055} (we do not have the corresponding suitable R band image to perform continuum subtraction making the H$\alpha$ fluxes too unreliable to be used here) and \textcolor{black}{NGC~5457} is responsible for this difference. Among those regions, 213 (respectively 184) are detected with $S/N>3$ (resp. $S/N>5$). It shows that the star forming regions are young and have recently formed stars. Indeed, the H$\alpha$ equivalent width is a good tracer of the age of a single star forming episode for the first $\sim10-15\times10^6$~years. We will see in section \ref{ssec:eval-age} that the $U-B$ color is also a good tracer of the age.

\subsection{Estimation of the extinction}

In order to estimate attenuation values for our regions, we have made use of the 24~$\mu$m and the H$\alpha$ images. Indeed \cite{calzetti2007a} showed that there is a very good correlation between the extinction corrected Pa$\alpha$ and a combination of the 24~$\mu$m and dust obscurated H$\alpha$ leading to an estimate of the intrinsic H$\alpha$ luminosity:

\begin{equation}
 L\left(\textrm{H}\alpha\right)_{intrinsic}=L\left(\textrm{H}\alpha\right)_{observed}+0.031\times L\left(24\right).
\end{equation}
This permits to derive the extinction of the gas:
\begin{equation}
 A_V^{gas}=3.06\times\log\left[1+\frac{0.031\times L\left(24\right)}{L\left(\textrm{H}\alpha\right)_{observed}}\right].
\end{equation}
As the extinction of the stars is smaller, we assume here that $A_V=0.44\times A_V^{gas}$ \citep{calzetti2000a}. We obtain an extinction ranging from $A_V=0.06$ to $A_V=1.28$ with a mean $\left<A_V\right>=0.44\pm0.28$.

Note that as this method to estimate the extinction is dependent on the H$\alpha$, \textcolor{blue}{NGC~5055} and \textcolor{black}{NGC~5457} fluxes cannot be corrected for the extinction and are therefore excluded from any subsequent analysis involving extinction correction.

\section{Modeling}
\label{sec:modeling}

\subsection{Spectral synthesis}

To model the SED of individual star forming regions, we have used the {\sc P\'egase ii} evolutionary spectral synthesis code \citep{fioc1997a}. We have assumed an instantaneous starburst having an age ranging from $1\times10^6$ years to $300\times10^6$ years with a constant metallicity of $Z=0.02$ and a \citet{kroupa2001a} initial mass function from 0.1~M$_\sun$ to 120~M$_\sun$. The impact of assuming solar metallicity in the models on our results will be examined in section \ref{ssec:eval-age}.

Fluxes and colors are determined convolving the generated SED -- taking into account the nebular emission lines as well as the nebular continuum which is important for young regions \citep{anders2003a} -- with the spectral response function of the filters for the FUV, NUV, U and B bands. The infrared luminosity is obtained calculating the absorbed flux by the dust. See the section hereafter for details about the dust models.

\subsection{Dust models}

We use different dust models in order to probe the effect of dust-stars geometry: 1) the starburst attenuation curve \citep{calzetti2000a}: classically used in deep surveys with the IRX-$\beta$ relation, 2) the foreground dust screen Small Magellanic Cloud attenuation curve \citep{gordon2003a}, 3) the mixed stars and dust Small Magellanic Cloud attenuation curve using the relation: $I_{\textrm{observed}}/I_{\textrm{intrinsic}}=0.921\times\left(1-10^{-0.4A\left(\lambda\right)}\right)/A\left(\lambda\right)$ adapted from \cite{natta1984a}.

The dust extinction is applied to the unextinguished SED with $A_V$ ranging from 0 to 3 by steps of 0.1. In this paper, conventionally, $A_V$ designates the extinction of the stars. The extinction of the gas is noted $A_V^{gas}$. The nebular emission accounts for a small fraction of the total flux in the broad bands, and the same extinction law has been applied to the stellar and the nebular emissions in these cases.

A comparison with the models generated by \citet{calzetti2005a} using {\sc Starburst99} shows that the results are nearly identical.

\section{The age as the ``second parameter''?\label{sec:age}}

In order to probe the driving parameters for the spread of the normal star-forming galaxies in the IRX-$\beta$ diagram, we need to derive the following quantities for our galaxies: the ultraviolet slope, the total infrared luminosity (section \ref{ssec:eval-IR}), which will be used to derive the FIR/FUV ratio,  and estimators for the age of the stellar populations. In the present work, we use two estimators: the $U-B$ color and the equivalent width of H$\alpha$ emission line (section \ref{ssec:eval-age}).

\subsection{Estimation of the bolometric infrared luminosity\label{ssec:eval-IR}}

In order to determine the total infrared emission of the selected star forming region we use the Spitzer IRAC 8.0~$\mu$m and MIPS 24~$\mu$m emission. While the exploitation of the MIPS 70~$\mu$m and 160~$\mu$m bands would yield better results, their resolution (17\arcsec~and 38\arcsec) is too low to separate individual star forming regions. For this study we estimate the total infrared emission using the \citet{calzetti2005a} relation that was derived in a preliminary study on \textcolor{magenta}{NGC~5194}:

\begin{equation}
 \log L\left(IR\right)=\log\left[L\left(24\right)\right]+0.908+0.793\log\left[L_\nu\left(8\right)/L_\nu\left(24\right)\right].
\end{equation}

The scatter around the relation within a galaxy is about 40\% \citep{calzetti2005a}, which is the accuracy level we can expect for the determination of the total infrared flux of individual star forming regions. \cite{perez2006a} also derived a similar relation. Practically, the difference between the \cite{calzetti2005a} and the \cite{perez2006a} IRX estimates for the selected star forming regions is $-0.07\pm0.03$ dex which is only a few percent of the total range and does not alter the results presented in this study.

\subsection{Stellar population age estimators\label{ssec:eval-age}}

\begin{figure*}[!ht]
\includegraphics[width=\columnwidth]{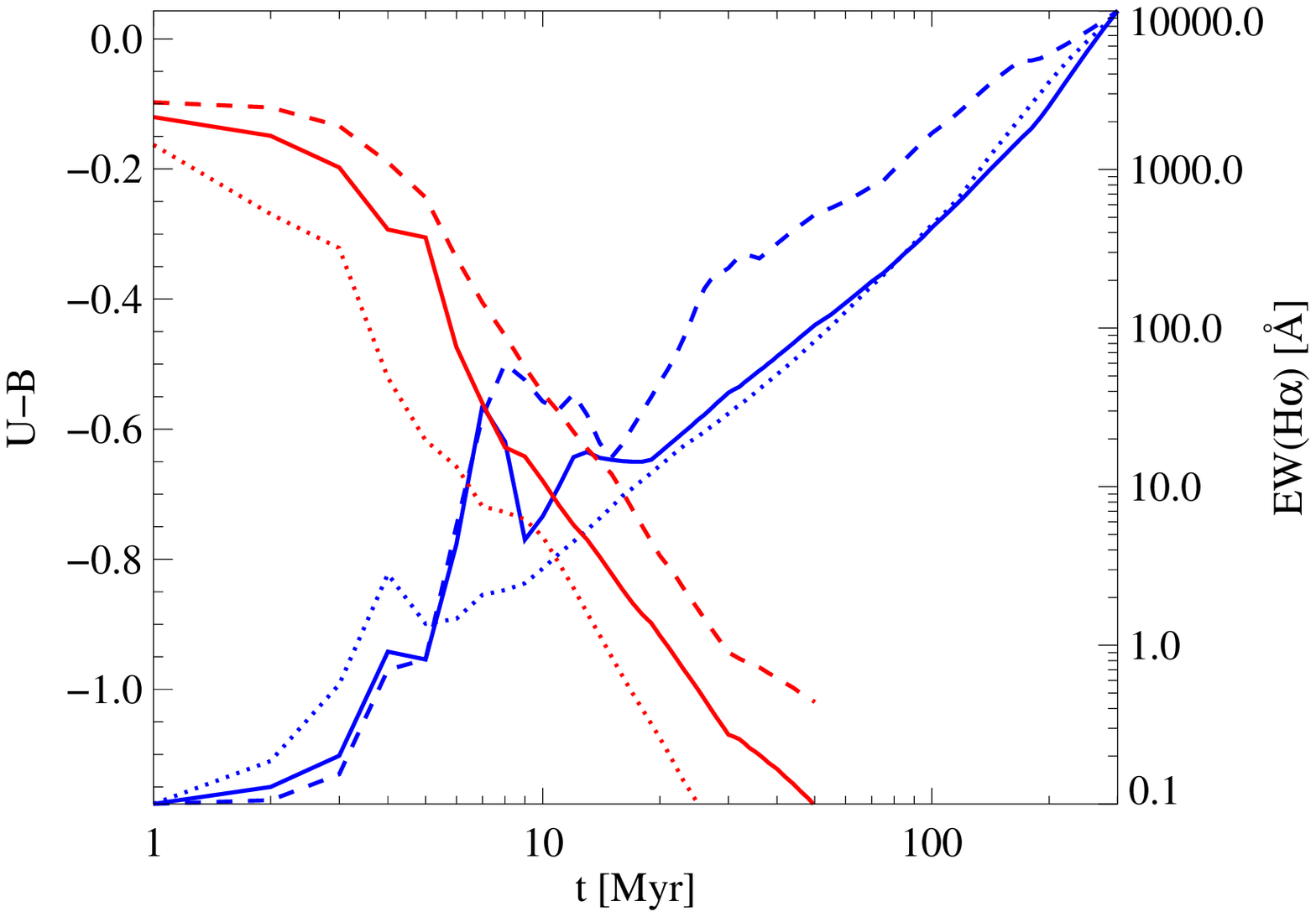}
\includegraphics[width=\columnwidth]{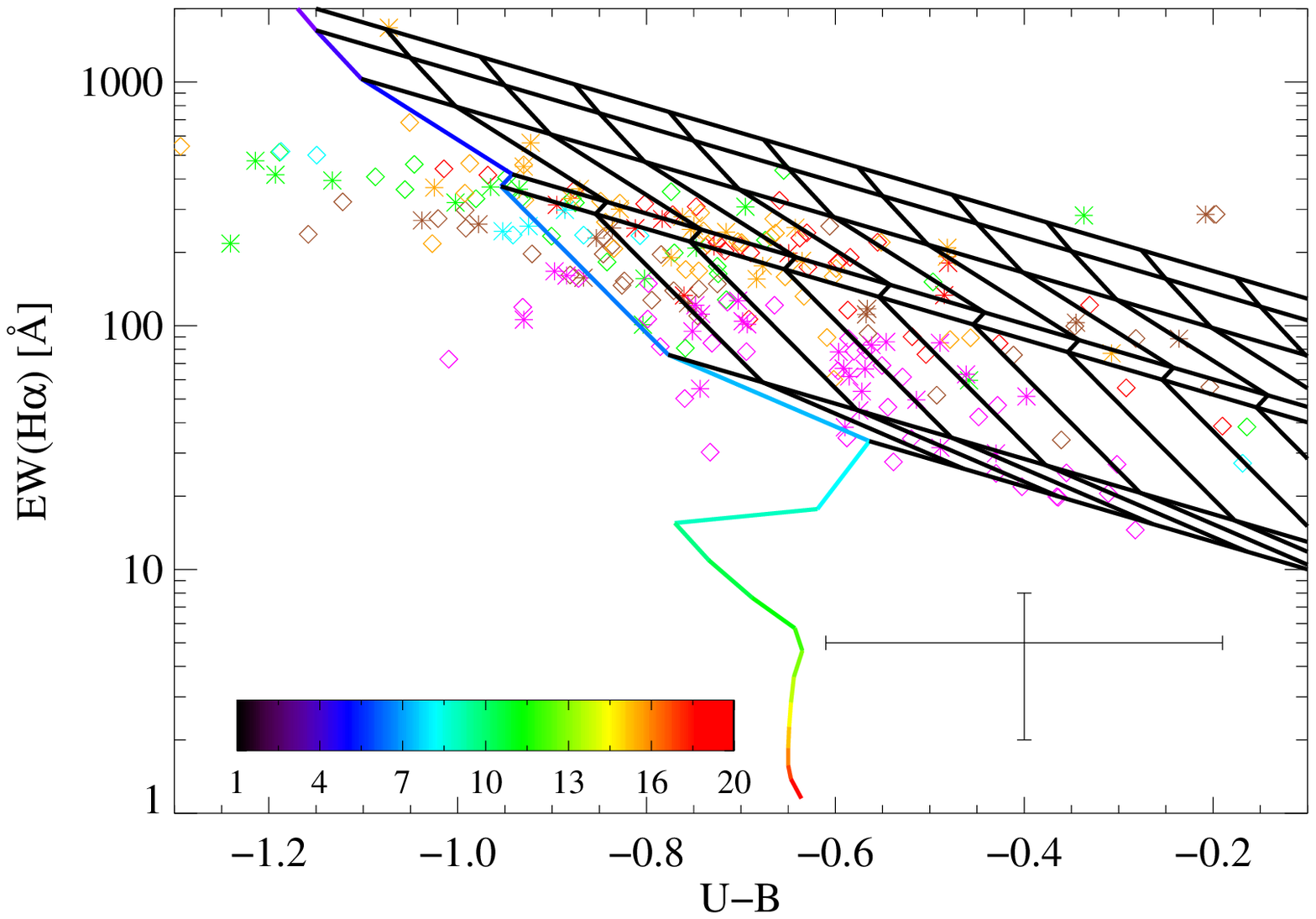}
\caption{Left: $U-B$ color (blue line) and the H$\alpha$ equivalent width (red line) versus the time elapsed since an extinction-free instantaneous starburst with a \cite{kroupa2001a} IMF and a metallicity $Z=0.004$ (dashed line), $Z=0.02$ (solid line) and $Z=0.01$ (dotted line). Note that practically measurements of equivalent widths smaller than a few \AA\ are uncertain. Right: equivalent width of the H$\alpha$ line defined as $L\left(H\alpha\right)/L_\lambda\left(R\right)$ in units of \AA. The colors are the same as defined in Figure~\ref{fig:24_vs_fuv}. The solid colored line represents the track of an extinction-free instantaneous starburst with a \cite{kroupa2001a} IMF and a metallicity $Z=0.02$. The age ranges from $10^6$~years to $20\times10^6$~years by increments of $10^6$~years. It is indicated by the colorbar. The black solid lines departing from the colored solid line represent the effect of the extinction, assuming that $E\left(U-B\right)=0.24\times A_V$ mag and $EW\left(H\alpha\right)\propto10^{-0.4\times0.68\times A_V}$ mag which takes into account the differential extinction between the gas and the R band emitting population. Each intersecting line represents an increment of the extinction by 0.1 magnitude.\label{fig:UV-EW(Ha)}}
\end{figure*}

To test if the age is the ``second parameter'' we are looking for, an accurate determination of the stellar population ages is needed. Given the large number of star forming regions involved, a direct spectral observation of the age sensitive 4000~\AA\ break, as measured by the $D_{4000}$ parameter are not available for the majority of regions in our sample. We have decided to exploit the $U-B$ color as a proxy to the 4000~\AA\ break. Indeed the U band is centered around 360~nm while the B band is centered around 440~nm, on each side of the break. A simple {\sc P\'egase II} model using a \cite{kroupa2001a} IMF from 0.1~M$_\sun$ to 120~M$_\sun$ and a metallicity of Z=0.02 shows that for an instantaneous burst, the $U-B$ color ranges from $-1.18$ after $1\times10^6$ years to $0.04$ after $300\times10^6$ years (Figure~\ref{fig:UV-EW(Ha)} left).

While the $D_{4000}$ parameter is essentially extinction free thanks to the narrow wavelength range probed, extinction is strong at short wavelengths, inducing an important uncertainty regarding the broadband U and B fluxes. However this problem is mitigated when using the $U-B$ color. Indeed, when averaged on a number of dust geometries and extinction curves, an extinction of $A_V=1$ yields only $E\left(U-B\right)\sim0.1\times A_V$ mag. We will see in section \ref{ssec:irx-beta} that the stellar extinction -- assuming $A_V^{stars}=0.44\times A_V^{gas}$ -- ranges from $A_V=0.06$ and $A_V=1.28$, with a mean extinction of 0.44 magnitude, which yields an average uncertainty of 0.04 magnitude (0.1 magnitude if we assume a starburst attenuation law).

Young star forming regions are the main source of H$\alpha$ emission in star forming galaxies. As the population ages, the H$\alpha$ drops dramatically as stars emitting ionizing photons have a life expectancy typically under $20\times10^6$ years. A standard way to estimate the age is the equivalent width of the H$\alpha$ emission line, which decreases as the most massive stars die. The exact estimation would necessitate spectroscopic observations. Indeed, even if the R band emission is not constituted only of the continuum -- it is ``polluted'' by the H$\alpha$ and [NII] lines and the nebular emission in general -- it is the best possible approximation with the available data. Another advantage is to limit the differential extinction as the H$\alpha$ line and the R band central wavelength are close.

A comparison of the evolution of $U-B$ and H$\alpha$ equivalent width age tracers against time is plotted in Figure~\ref{fig:UV-EW(Ha)} left and they are compared with observations in Figure~\ref{fig:UV-EW(Ha)} right.

We see in Figure~\ref{fig:UV-EW(Ha)} right that for all star forming regions, the H$\alpha$ equivalent width is reasonably high, ranging typically from $\sim15$~\AA\ to $\sim600$~\AA. This indicates a young age, typically less than $10\times10^6$~years. The $U-B$ color ranges typically from $-1.3$ to $-0.1$. There is a clear correlation between the H$\alpha$ equivalent width and the $U-B$ color with a Pearson correlation coefficient $r=-0.52$. \textcolor{magenta}{NGC~5194} star forming regions are significantly offset from the mean correlation, having a much lower H$\alpha$ equivalent width for their $U-B$ color (or having a much bluer $U-B$ color for their H$\alpha$ equivalent width). The instantaneous burst model with a \cite{kroupa2001a} IMF and a metallicity of $Z=0.02$ can reproduce most star forming regions that have a $U-B$ color redder than $-0.9$ assuming an extinction such as $E\left(U-B\right)=0.24\times A_V$ mag and $EW\left(H\alpha\right)\propto10^{-0.4\times0.68\times A_V}$ with $0.04<A_V<1.28$. This shows that while the $U-B$ color is only weakly sensitive to the extinction, the inclusion of extinction effects is nevertheless needed to reproduce most of the observations. However for the bluest $U-B$ star forming regions, the model significantly overpredicts the observations of the H$\alpha$ equivalent width. One possible explanation is an insufficient background removal in the R band image within each region, artificially decreasing the $EW\left(H\alpha\right)$ values for blue $U-B$ colors. An alternative explanation is a strong contamination of the R band by the H$\alpha$ and [NII] lines which would lead to significantly overestimate the continuum in the R band. However, the mean EW(H$\alpha$) is $195\pm193$~\AA\ and the typical width of a R band filter is $\sim1500$~\AA. This means the contamination is at most about 25\%, insufficient to reproduce the observations. In section \ref{ssec:irx-beta} we will see that the characteristics of most of the star forming can be reproduced by a simple model in the IRX-$\beta$ diagram. An important fraction of them however should not present any significant H$\alpha$ emission -- 73\% (resp. 60\%) should be older than $12\times10^6$ years (resp. $20\times10^6$ years) -- if the underlying hypotheses (the star formation history can be modeled as an instantaneous burst and the trend observed as a function of the perpendicular distance is an age effect) are correct.

\begin{figure}[!ht]
\includegraphics[width=\columnwidth]{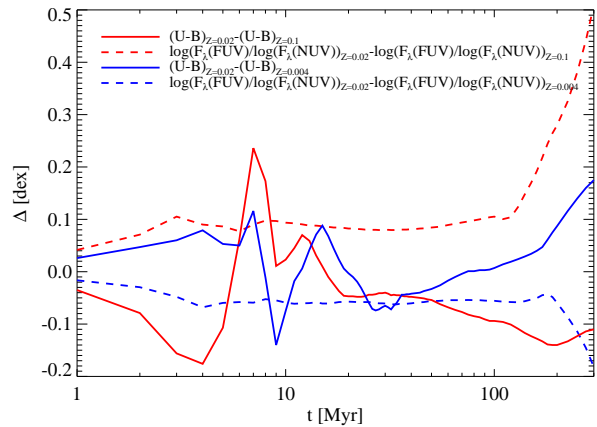}
\caption{Effect of the metallicity on the $U-B$ and ultraviolet colors. The solid lines represent the $U-B$ colors while the dashed ones represent the $\log\left(F_\lambda\left(FUV\right)/F_\lambda\left(NUV\right)\right)$ colors. The red lines give the difference between $Z=0.02$ and $Z=0.1$ while the the blue ones give the difference between $Z=0.02$ and $Z=0.004$. For an ionizing population $\left(U-B\right)_{Z=0.1}-\left(U-B\right)_{Z=0.004}$ (resp. $\log\left(F_\lambda\left(FUV\right)/F_\lambda\left(NUV\right)\right)_{Z=0.1}-\log\left(F_\lambda\left(FUV\right)/F_\lambda\left(NUV\right)\right)_{Z=0.004}$) reaches a maximum of $0.26$ (resp. $0.16$).\label{fig:UmB-metal}}
\end{figure}

A parameter that may affect the age estimate is the metallicity. Indeed, higher metallicity stars have a significantly redder $U-B$ color due to the presence of a high number of metallic absorption lines in their atmosphere. An illustration of this effect on the $U-B$ and the ultraviolet colors is presented in Figure~\ref{fig:UmB-metal}.

While the metallicity has an influence, it is small as the oxygen abundances are typically in the relatively narrow range $12+\log O/H=$8.6--9.2, except NGC~5457 which outermost regions have a metallicity down to 8.2. As a reference, \cite{asplund2005a} determined the solar metallicity as $Z=0.0122$ or $12+\log O/H=8.66$.

Another uncertainty is due to the synthesis of the U-band luminosities. Indeed, even in simple $U-B$, $B-V$ color-color diagrams, the model tracks are not consistent with integrated photometry of dwarf galaxies \citep[][for more details]{lee2006a}.

\subsection{IRX-$\beta$ diagram\label{ssec:irx-beta}}

\begin{figure*}[!ht]
 \includegraphics[width=\columnwidth]{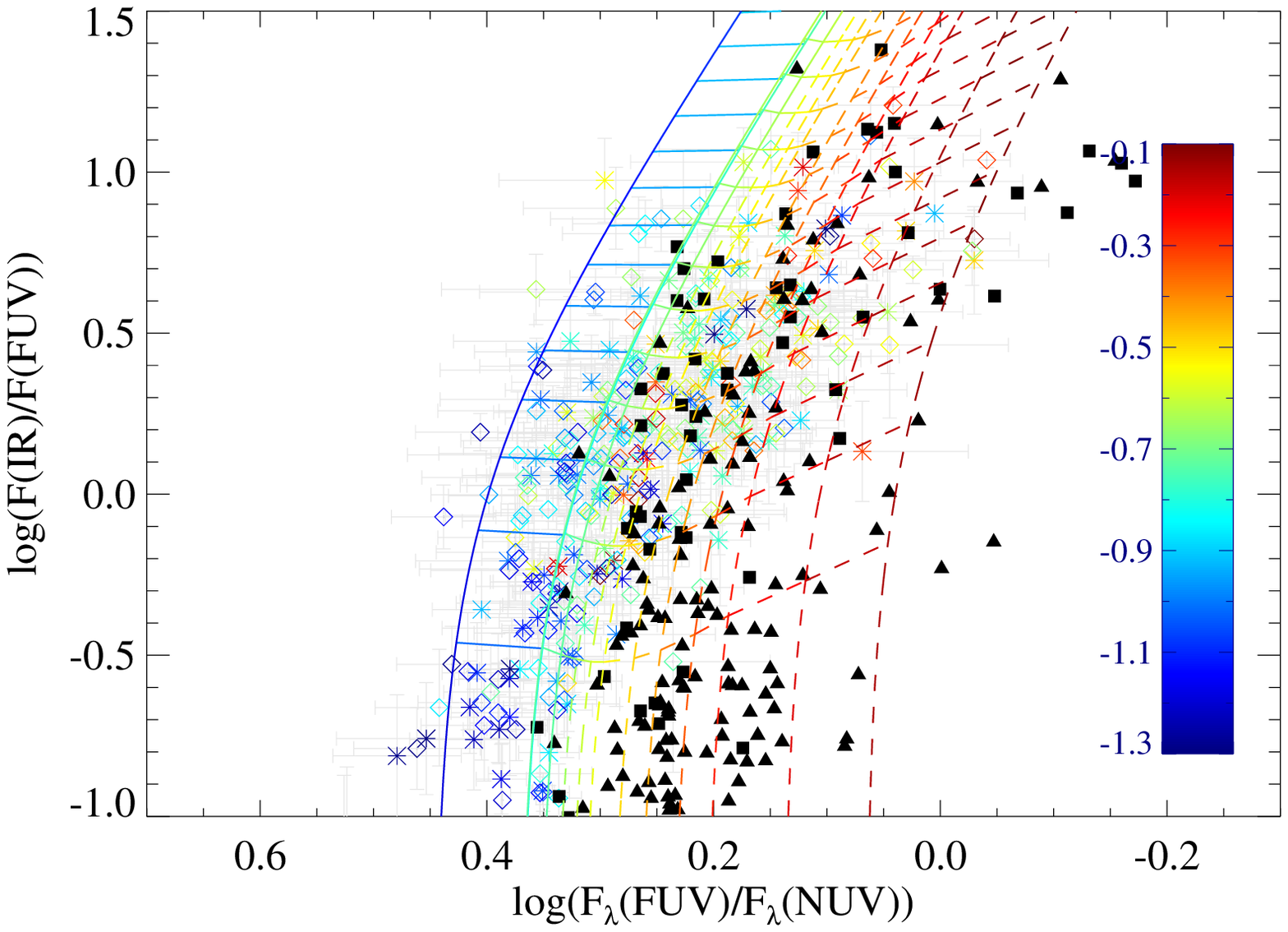}
 \includegraphics[width=\columnwidth]{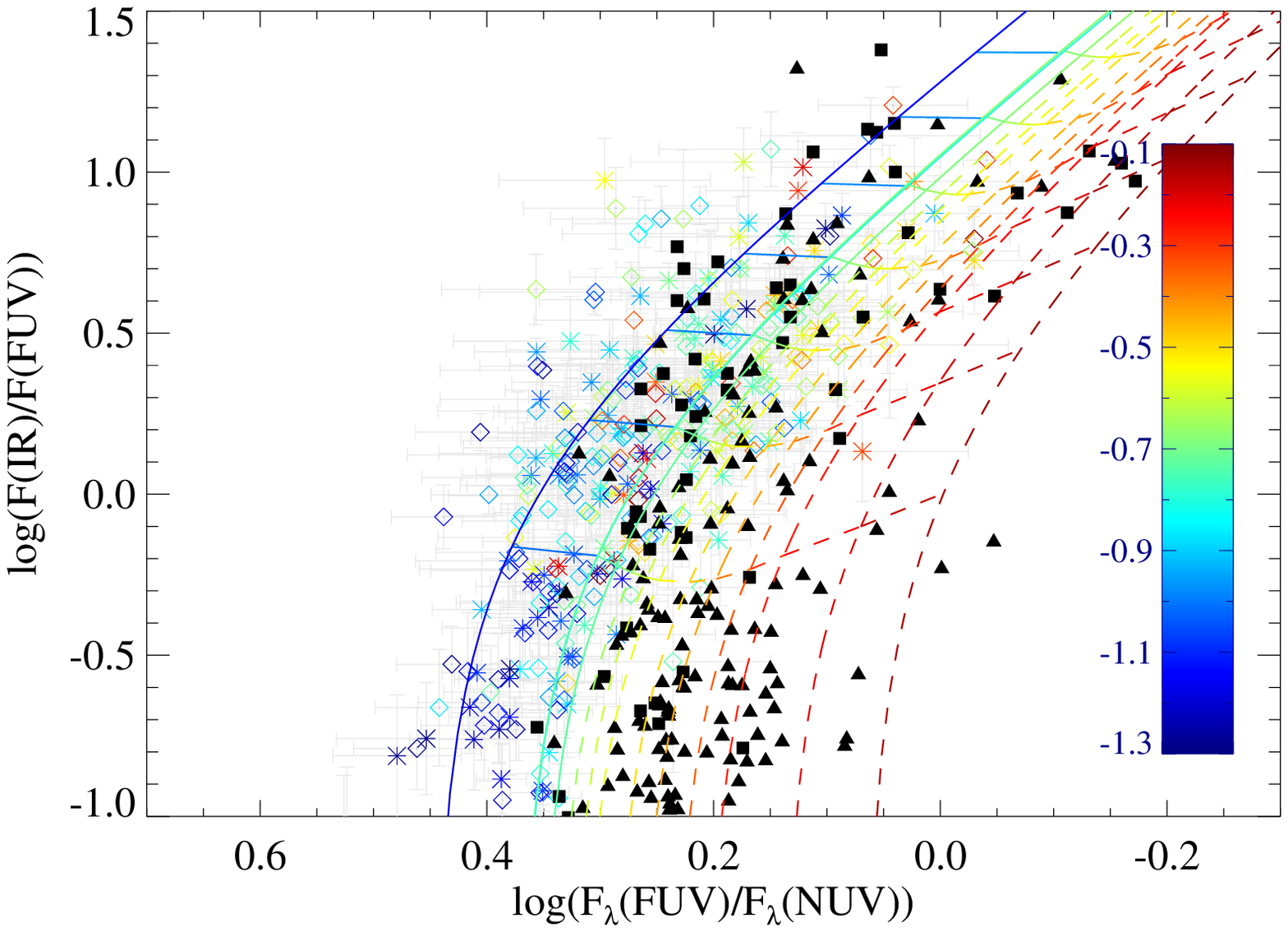}
 \caption{IRX-$\beta$ diagram of individual star forming regions. The regions represented by a diamond (resp. star) are FUV (resp. 24~$\mu$m) selected. The color of the symbols indicates the actual $U-B$ color (i.e. the age) following the scale defined by the color bar on the right corner. The error bars are plotted in light gray. The grid is the model described in section \ref{sec:modeling} using a starburst attenuation curve (left) and the SMC extinction curve (right). The color of each vertex corresponds to the expected $U-B$ color. The axis predominantly horizontal (resp. vertical) samples the age: $1\times10^6$, $6\times10^6$, $12\times10^6$, $20\times10^6$, $30\times10^6$, $40\times10^6$, $50\times10^6$, $75\times10^6$, $100\times10^6$, $150\times10^6$, $200\times10^6$, $250\times10^6$ and $300\times10^6$ years (resp. extinction from $A_V=0.1$ to $A_V=3.0$, the first horizontal line representing an extinction of 0.1 magnitude.). The solid lines represent an ionizing population while the dashed lines represent a non-ionizing population. The model representing the origin of the perpendicular distance ($d=0$) corresponds to an age of $6\times10^6$ years. We see on the grid that the age has a much stronger effect on the $U-B$ color than the extinction. The black squares represent SINGS \citep{kennicutt2003a} galaxies and the black triangles LVL ones (Dale et al., 2009).\label{fig:irx-beta}}
\end{figure*}

In Figure~\ref{fig:irx-beta} we present the IRX-$\beta$ diagram of the selected star forming regions compared with the model presented in section \ref{sec:modeling}.

Several results have to be noted. We see that most of the star forming regions are in general agreement with the models though a number of regions have a conspicuously blue UV color or a particularly low IRX. We also see that the selected star forming regions describe a clear envelope mostly circumscribed in its upper part by the modeled starburst IRX-$\beta$ relation of a very young starburst. We retrieve here the well known result that quiescent star forming galaxies lay under the starburst IRX-$\beta$ curve.

A visual inspection shows an expected qualitative trend to have younger regions where the UV color is the bluest while on the right part there are predominantly older regions which have a redder UV color. Interestingly the younger regions have a trend closer to the modeled IRX-$\beta$ starburst relation. One explanation could be that it marks the locus of both constant star formation and young stellar regions. Indeed, in addition to their young age, the dust-stars geometry is likely to be similar to that of starburst galaxies (star forming regions inside a dust shell). Another explanation could be that those regions are still relatively enshrouded in their birth cloud. However, as shown by \cite{prescott2007a}, only 3\% of regions are completely enshrouded in dust.

\subsection{Age sensitive tracers as a function of the perpendicular distance\label{ssec:perp-dist}}

Here we quantify whether age may be the parameter accounting for the deviations of our regions from the starburst attenuation curve in the IR-$\beta$ diagram.

A standard tool to evaluate the distance of a star forming region (or a galaxy as a whole) from the standard starburst attenuation law is the perpendicular distance in the IRX-$\beta$ diagram \citep[][for instance]{kong2004a}. The distance is simply defined as the euclidean distance:

\begin{equation}\label{eqn:dist}
 d=\sqrt{\left(x-x_{EL}\right)^2+\left(y-y_{EL}\right)^2},
\end{equation}
where $x$ (resp. $x_{EL}$) and $y$ (resp. $y_{EL}$) are the $\beta$ and IRX values of the individual star forming regions (resp. of the extinction law locus that is the closest to the observed value) and $d$ is the dimensionless distance in the IRX-$\beta$ plot. The starburst and SMC IRX-$\beta$ relations are obtained from the model presented above. The SMC extinction law will be studied more in detail in section \ref{sec:extinc-laws}. We assume an age of $6\times10^6$ years which gives an intrinsic $\log F_\lambda\left(FUV\right)/F_\lambda\left(NUV\right)=0.37$. The IRX-$\beta$ relation is obtained varying the extinction applied to this population. Following the \cite{kong2004a} convention, star forming regions which have a higher (resp. lower) than expected IRX given their UV slope will have $d>0$ (resp. $d<0$).

\begin{figure*}[!ht]
 \includegraphics[width=\columnwidth]{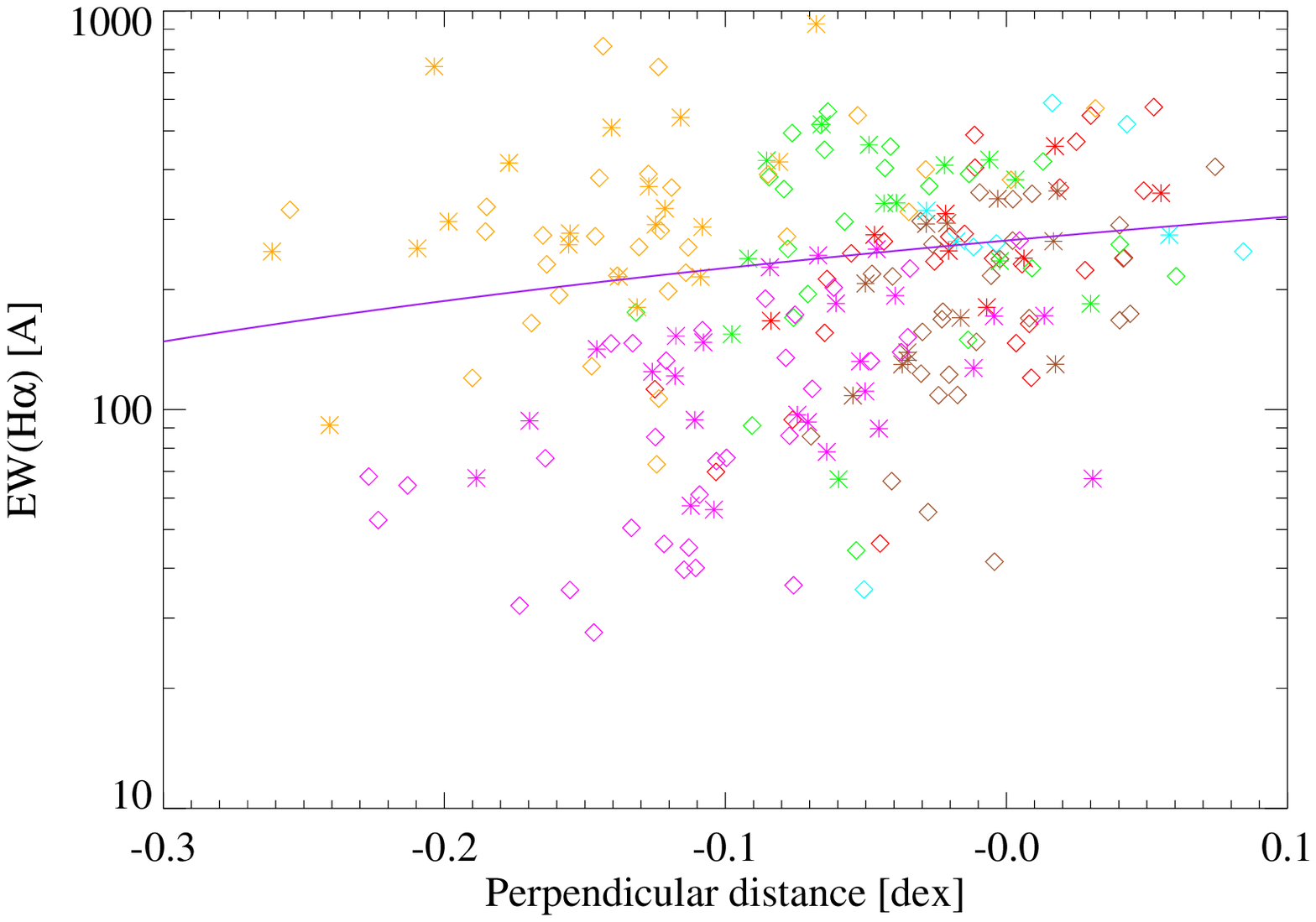}
 \includegraphics[width=\columnwidth]{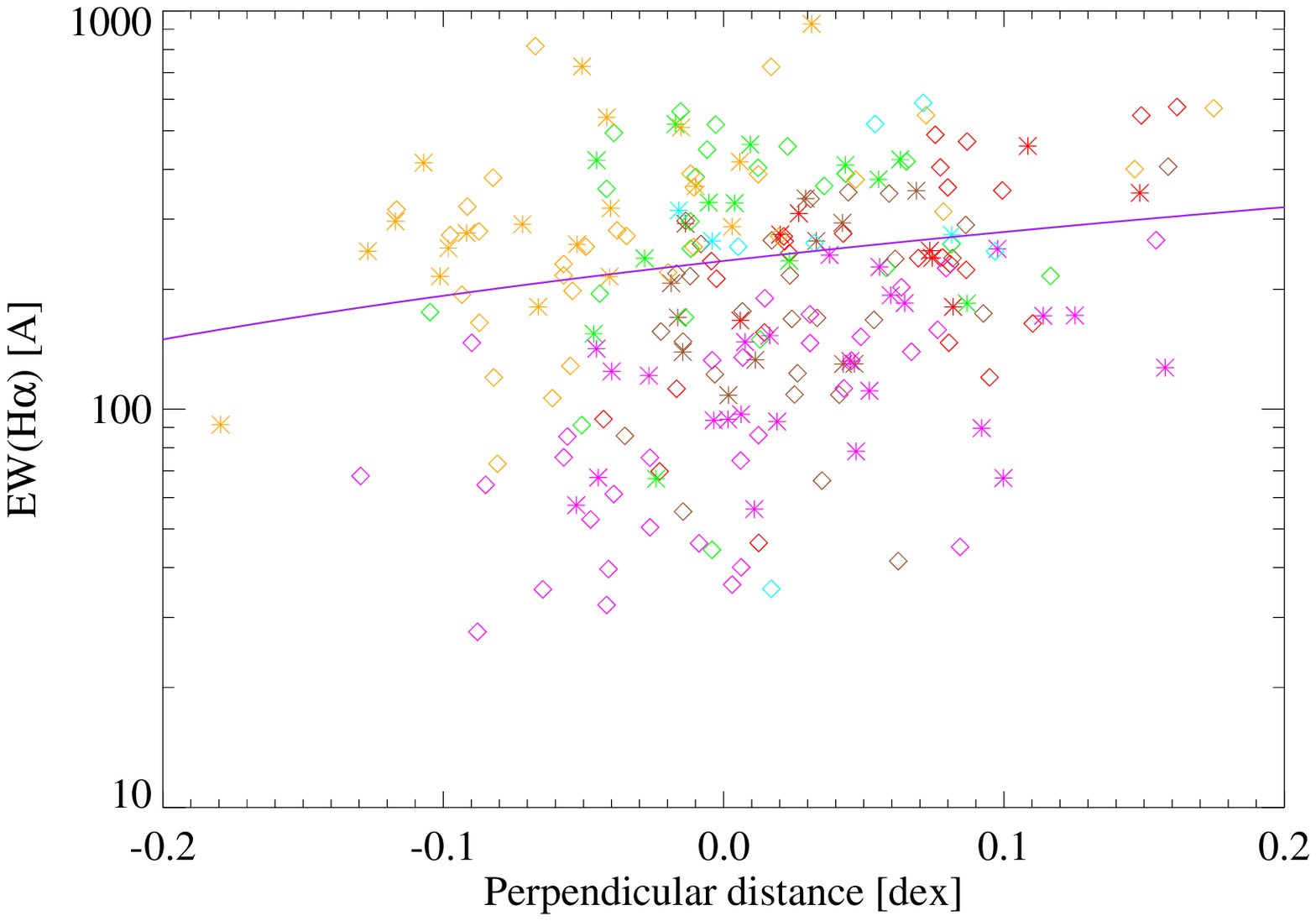}
 \caption{Differential extinction corrected H$\alpha$ equivalent width, as defined by the ratio of the H$\alpha$ over R band flux density, as a function of the perpendicular distance (as defined in equation \ref{eqn:dist}, see section \ref{ssec:perp-dist}) between the starburst (left) and SMC (right) laws and the individual star-forming regions in the IRX-$\beta$ diagram. The purple line is the fit of the equivalent width as a function of the perpendicular distance: $EW\left(H\alpha\right)=264\pm13+\left(385\pm144\right)\times d$ (starburst) and $EW\left(H\alpha\right)=234\pm10+\left(410\pm157\right)\times d$ (SMC).\label{fig:dist-EW-Ha}}
\end{figure*}

In Figure \ref{fig:dist-EW-Ha} we plot the equivalent width as the ratio between the H$\alpha$ flux and the R band flux density corrected for the extinction using the formula in section \ref{ssec:eval-age} against the perpendicular distance.

As expected the mean H$\alpha$ equivalent width increases as a function of the perpendicular distance with a slope of $385$ and a large $1\sigma$ scatter around the linear fit: $150$\AA. The perpendicular distance and the H$\alpha$ equivalent width are weakly correlated with a Pearson correlation coefficient $r=0.18$. The most surprising result is that even at large negative perpendicular distances, a significant H$\alpha$ emission is detected. This result shows that the age as measured by the H$\alpha$ equivalent width is not the ``second parameter'' we are seeking. From the point of view of star formation, it could mean that the assumption of an instantaneous starburst is not valid. We will explore if variations of the star formation history can explain the observations in section \ref{ssec:disc-dilution}.

\begin{figure*}[!ht]
 \includegraphics[width=\columnwidth]{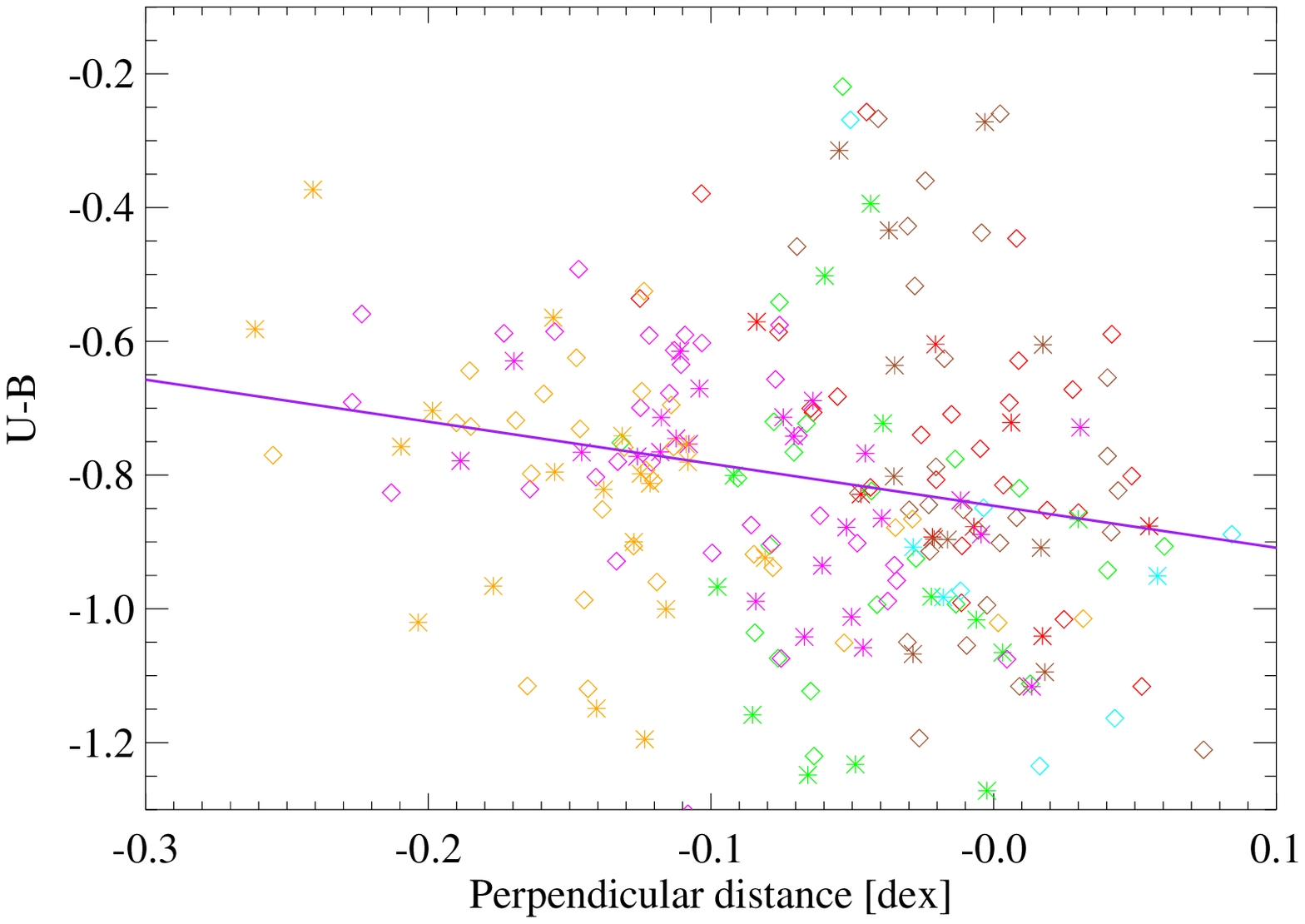}
 \includegraphics[width=\columnwidth]{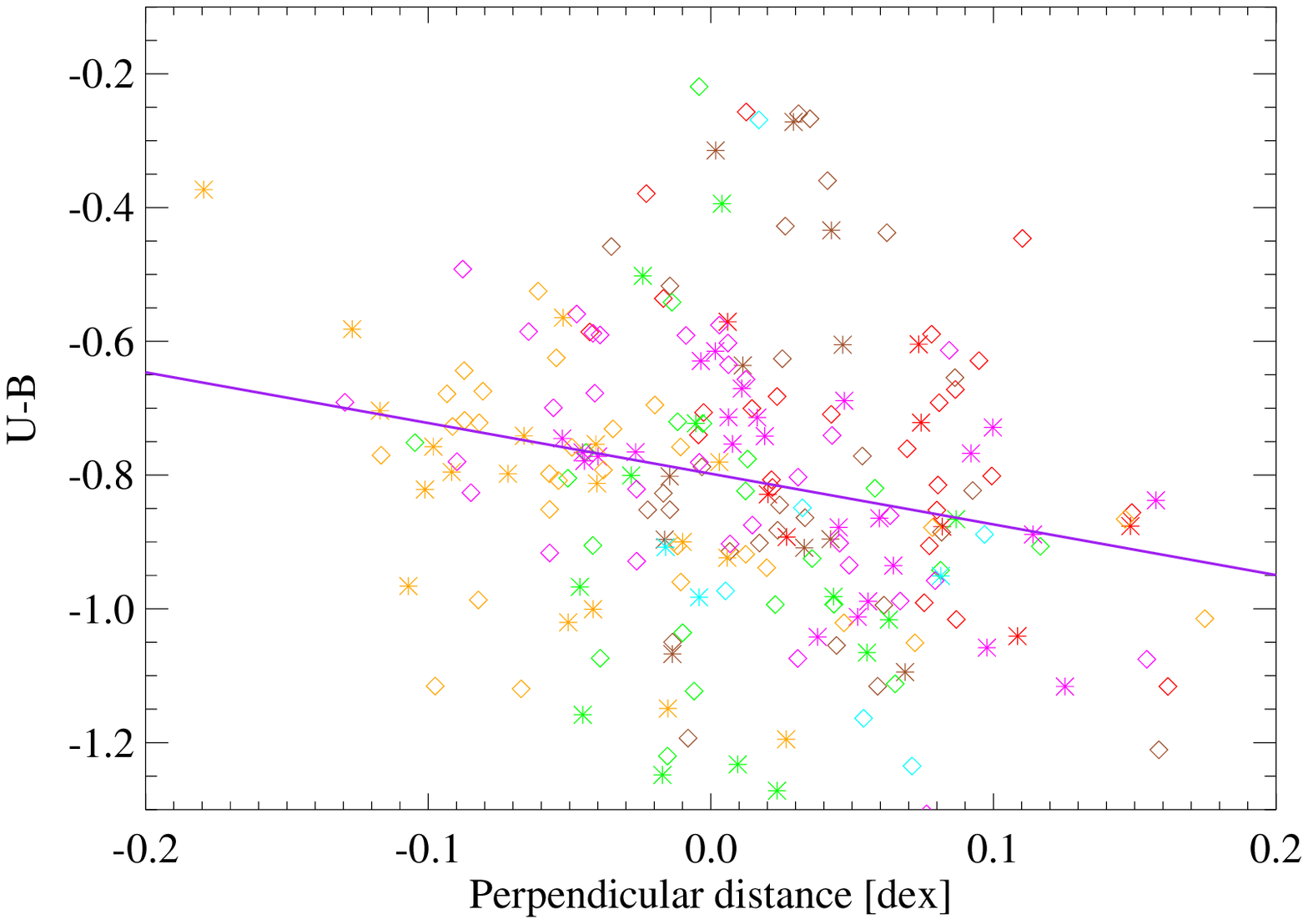}
 \caption{Extinction corrected $U-B$ color as a function of the perpendicular distance (as defined in equation \ref{eqn:dist}, see section \ref{ssec:perp-dist}) between the starburst (left) and SMC (right) laws and the individual star-forming regions in the IRX-$\beta$ diagram. The purple line is the linear fit of the extinction corrected $U-B$ color as a function of the perpendicular distance: $U-B=-0.85\pm0.02-\left(0.62\pm0.20\right)\times d$ (starburst) and $U-B=-0.80\pm0.01-\left(0.74\pm0.22\right)\times d$ (SMC).\label{fig:dist-U-B-corrected}}
\end{figure*}

In Figure~\ref{fig:dist-U-B-corrected} the extinction corrected $U-B$ -- assuming a starburst attenuation law: $E\left(U-B\right)=0.24\times A_V$~mag -- is plotted as a function of the perpendicular distance.

While a trend of the $U-B$ color as a function of the perpendicular distance could be expected, it is particularly weak, with a slope of only $-0.62\pm0.20$. The perpendicular distance and the $U-B$ color are correlated with a Pearson correlation coefficient $r=-0.17$. The weakness of the trend combined with the important scatter hints that another physical phenomenon than the age plays a much more important role.

%

We have shown here that the age, which was a prime candidate to explain the deviation from the starburst attenuation law in the IRX-$\beta$ diagram does not seem to play a major role. Indeed, using two different age tracers, the $U-B$ color and the H$\alpha$ line equivalent width, the large dispersion around the relation shows that the influence of the age is at most minimal under the current assumptions and for the timescales probed by those two tracers.

\subsection{Influence of the star formation history and dilution by an older stellar population\label{ssec:disc-dilution}}

We have shown that an instantaneous starburst cannot reproduce simultaneously the large perpendicular distances and a non negligible H$\alpha$ emission. In particular, there are a number of regions in our sample that are too red in UV colors. We now consider the possibility that the stellar populations in those star forming regions are more complex than a simple instantaneous burst description. We thus introduce a model of the stellar population of a galactic disk that considers an exponentially decreasing SFR of the form:

\begin{equation}\label{eqm:sfr-exp}
 SFR\left(t\right)=a\times\exp\left(-t/\tau\right),\label{eq:exp}
\end{equation}
where $a$ is an arbitrary normalization constant in units of M$_\sun$.yr$^{-1}$, $t$ is the time in years and $\tau$ is the time constant in years.

First of all we want to see if an exponentially decreasing star formation rate is able to reproduce the observations. To take into account different cases of star formation history we test 3 models which differ by the ratio $SFR\left(12\times10^9\right)/SFR\left(0\right)$. $SFR\left(0\right)$ is the SFR at $t=0$~year and $SFR\left(12\times10^9\right)$ is the SFR at $t=12\times10^9$~years, both from equation \ref{eq:exp}. The values of this ratio are $1/10$, $1/100$ and $1/1000$ (which correspond to a birthrate parameter of $2.6\times10^{-1}$, $4.7\times10^{-2}$ and $6.9\times10^{-3}$ and a timescale of $5.2\times10^9$~years, $2.6\times10^9$ years and $1.73\times10^9$ years respectively). It turns out that this star formation history can be readily excluded as it cannot reproduce at all the observations. The reason is that even for $SFR\left(12\times10^9\right)/SFR\left(0\right)=1/1000$, the current star formation rate is high enough to overwhelm ultraviolet emission from older stellar populations.

\begin{figure}[!ht]
 \includegraphics[width=\columnwidth]{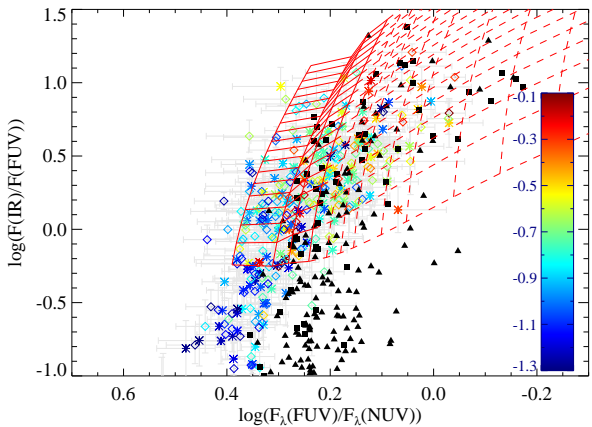}
 \caption{Same as Figure~\ref{fig:irx-beta} with a model constituted of an exponentially decreasing burst during $11.95\times10^{12}$ years and an instantaneous starburst of a mass of $M_{stars}/300$ at $t=12\times10^9$.\label{fig:irx-beta-model-exp-plus-burst}}
\end{figure}

A more realistic model for a galactic star forming region is a small burst on top of the galactic stellar population. To model the galactic stellar population we use an exponentially decreasing star formation rate as described in equation \ref{eqm:sfr-exp} for $0\le t\le11.95\times10^9$ years, no star formation for $11.95\times10^9<t<12\times10^9$ years and an instantaneous burst at $t=12\times10^9$ years. The rationale is that in the area of the star forming region, due to the diffusion of the stars over time, the stellar population is made of stars born in different regions of the galaxy subsequently mixed. However stars younger than a few $10\times10^6$ years are too young to have migrated long distances. In addition we assume that in the last $50\times10^6$ years, no star formation has taken place at the location of the current star forming region and no significant migration from nearby star forming regions has occurred. The mass of the instantaneous starburst at $t=12\times10^9$ years is defined as $M_{stars}/300$ following \cite{calzetti2005a}, were $M_{stars}$ is the mass of the stars constituting the background galactic population. The model is plotted in Figure~\ref{fig:irx-beta-model-exp-plus-burst} along with the observations in an IRX-$\beta$ diagram.

While this model reproduces a much larger range of ultraviolet colors as a whole, it still cannot explain the H$\alpha$ emission of such regions. Indeed, initially the instantaneous starburst emission dominates over the older population until it fades away after a few tens millions years, after which the older population dominates and is responsible of the redder ultraviolet colors compared to the model presented in section \ref{sec:modeling}.


While a continuous star formation does not account for the observations, we can take into account a more stochastic nature of the star formation history by considering a single, still UV emitting, older burst. Indeed, the presence of a significant old stellar population can have an influence on the UV color. Individual star forming regions with local backgrounds subtracted are less affected by this effect than integrated galaxies. Some of the observed regions are redder than can what can be reproduced by an ionizing population and a SMC extinction law (see Figure~\ref{fig:irx-beta} and section \ref{sec:extinc-laws}, the SMC extinction law is already extreme in that it has a very steep slope in the UV, this means that for a given IRX, it reddens more the UV color than any other extinction law considered in this study). That there are some regions which have an even redder color hints about the possibility of a dilution by older stellar populations. In order to test this, we have made a simple model assuming 2 instantaneous bursts, the one we are currently observing and another one which is $300\times10^6$ years old. The star formation history is undoubtedly more complex but cannot be precisely retrieved with the current observations. However, NUV emitting populations have an life expectancy of about $10^9$ years. We take here a $300\times10^6$ years old instantaneous burst as an approximation of the real star formation history. The mass ratio of the more evolved stellar population to the younger one varies from 0 to 300. We have selected all star forming regions that have a redder UV color than a $20\times10^6$ years old starburst extincted with the SMC extinction law. We then have calculated what dilution would be the necessary to redden the UV color, if those regions were actually $20\times10^6$ years old to account for the fact that they still emit in H$\alpha$. There is a total of 67 star forming regions which have a UV color redder than a SMC-extincted $20\times10^6$ years old burst. The model shows that the mean mass ratio is $7.8\pm7.1$ with a median of 5. This is much smaller than the ratio for integrated galaxies which reach up to few hundreds. This corresponds to a mean flux ratio of $0.15$ in FUV and $0.29$ in NUV. Note however that 64\% of those star forming regions have flux uncertainties such that the observations could be reproduced without needing an older stellar population.

\begin{figure}[!ht]
 \includegraphics[width=\columnwidth]{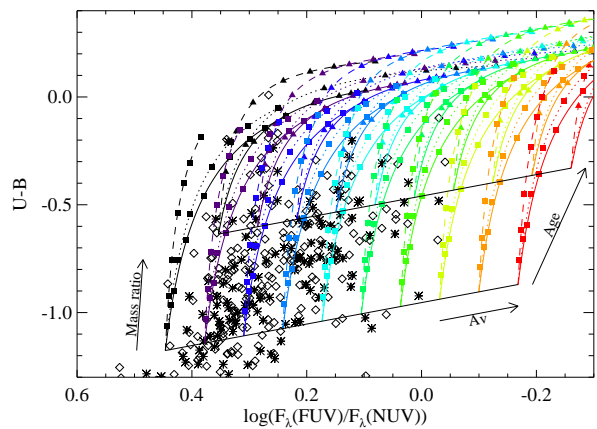}
 \caption{Influence of old stellar population on a $10^6$ and $20\times10^6$ years old, H$\alpha$ emitting, burst on the UV and $U-B$ colors. The straight solid lines represents the locus of a SMC-extincted $10^6$ (lower line) and $20\times10^6$ years old (upper line) bursts from $A_V=0.1$ to $A_V=1.0$. The color, solid, dotted and dashed lines trace the evolution of the colors as a function of the mass ratio for an old population of $300\times10^6$, $500\times10^6$ and $10^9$ years respectively. The successive filled squares (for the $20\times10^6$ years old population) and the triangles (for the $10^6$ years old population) indicate the mass ratio: 10, 25, 50, 100, 150, 200, 300.\label{fig:influence-dilution-colors}}
\end{figure}

The dilution by an old stellar population has an influence not only on the UV color but also simultaneously on the $U-B$ color as we can see in Figure~\ref{fig:influence-dilution-colors}.

We see that even a relatively small mass ratio can have a significant influence on the colors. The influence fades away as the older population gets older. Indeed, the intrinsic UV luminosity drops quickly with age. Hence, even if the older population gets redder, a higher mass ratio is needed to offset its smaller UV luminosity. On the other hand, the life expectancy of U and B emitting stars is much longer than those which predominate the UV luminosity. That is why an older population has significantly more influence on the $U-B$ color than on the UV color.

\section{Extinction laws}
\label{sec:extinc-laws}
We will explore here the effect of varying our assumptions on the dust attenuation law and dust geometry. We add to the starburst attenuation curve the case of an SMC extinction curve \citep{gordon2003a} for both foreground and mixed dust geometries. We will not consider curves like the Milky Way one, as \cite{calzetti2005a} showed that this curve cannot reproduce the observed trend of the data in the IRX-$\beta$ diagram, for any dust geometry. Conversely, the SMC extinction law, coupled with a foreground screen seems effective at reproducing part of the observed trend (Figure~\ref{fig:irx-beta}).

We observe that most of the star forming regions can be reproduced by one or more models but that no model can reproduce all observations. Because the SMC extinction law has a sharper rise compared to the starburst attenuation law, the UV color is more sensitive to the extinction in the former case than in the latter one, thus producing redder UV colors for equal FIR/UV ratios.

\begin{figure}[!ht]
 \includegraphics[width=\columnwidth]{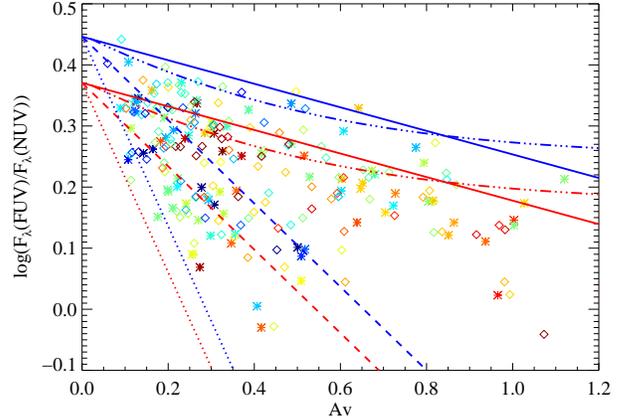}
 \caption{Ultraviolet color as a function of the extinction in magnitudes. Star forming regions are plotted with symbols and colors as defined in Figure~\ref{fig:irx-beta}. The starburst attenuation law is plotted with a solid line, SMC with a dashed line, SMC($A_V/0.44$) -- this is equivalent to $A_V=A_V^{gas}$, assuming that the relation $A_V=0.44\times A_V^{gas}$ is also valid in the case of a SMC attenuation law -- with a dotted line, and a mixed SMC law with a dash triple-dot line. The blue lines are for a $10^6$ years old burst and the red ones for a $12\times10^6$ years old burst.\label{fig:UV-color-Av-SB-SMC}}
\end{figure}

While the model using the SMC extinction law fails to reproduce the youngest regions that could be explained with a starburst attenuation curve it can however reproduce a number of UV red regions with a younger inferred age. The question of the reproducibility of the H$\alpha$ emission remains in fact open as most of the selected star forming regions are detected in H$\alpha$. In order to address this question we plot the UV colors that can be reproduced by H$\alpha$ emitting populations as a function of the extinction in Figure~\ref{fig:UV-color-Av-SB-SMC}. For each model we plot two lines, one for a $10^6$ year old starburst and one for an age of $12\times10^6$ years (the oldest age that can still produce a reasonably high number of ionizing photons).

We observe that as expected, a single extinction law cannot reproduce all of the observations. For instance, for a given extinction, in the case of the starburst attenuation law, the ultraviolet color is too blue. Regarding the SMC extinction law, the behavior is very different depending on the assumption made on the stellar extinction. If we assume $A_V=A_V^{gas}$, only the least extinguished star forming regions can be reproduced. Diminishing the extinction of the stars relative to the extinction of the gas, more star forming regions can be reproduced by the model, but only a fraction of the total number at a time (i.e. by each model).

So far we have always inspected the evolution of age tracers assuming the starburst attenuation law. As stated above, if the actual extinction law in quiescent star forming galaxies is different, it can emulate an age effect. As we have already specified, star forming regions in the IRX-$\beta$ diagram are bracketed by the starburst attenuation law and foreground SMC extinction law.

In Figure~\ref{fig:dist-EW-Ha} right, we plot the H$\alpha$ equivalent width versus the perpendicular distance from the SMC IRX-$\beta$ law. The trend between the H$\alpha$ equivalent width and the perpendicular distance is even weaker than in the case of a starburst attenuation curve: the slope is similar ($410$ versus $385$) as well as the Pearson correlation coefficient ($r=0.13$ in the SMC case and $r=0.18$ in the starburst one).

In Figure~\ref{fig:dist-U-B-corrected} right, we plot the $U-B$ age tracer versus the perpendicular distance from the SMC IRX-$\beta$ law. The trend between the $U-B$ and the perpendicular distance is approximately as weak as in the case of a starburst attenuation curve: the slope is similar ($-0.74$ versus $-0.62$) as well as the Pearson correlation coefficient ($r=-0.22$ versus $r=-0.17$).

The standard IRX-$\beta$ relation is based on the starburst attenuation law. However the physical conditions for such a law to be valid are not necessarily always met in the case of star forming regions. In this section we have investigated different extinction laws. We have shown that no single law is able to reproduce the observations. Conversely, a range of extinction laws between the starburst attenuation law and the foreground SMC  extinction law can explain the observations, in line with what was found by \cite{burgarella2005a}.

\section{Metallicity}
\label{sec:metallicity}
\begin{figure}[!ht]
 \includegraphics[width=\columnwidth]{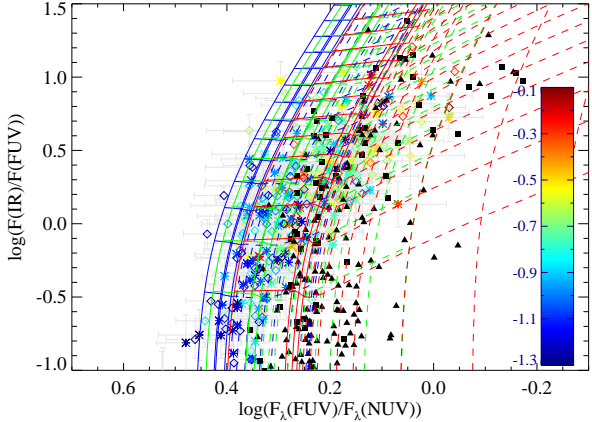}
 \caption{Same as Figure~\ref{fig:irx-beta}. The blue grid represents the model with \textcolor{blue}{$Z=0.004$}, the green one with \textcolor{green}{$Z=0.02$} and the red one with \textcolor{red}{$Z=0.1$}.\label{fig:irx-beta-grids-3-metallicities}}
\end{figure}

The metallicity can significantly affect the ultraviolet and $U-B$ colors as we have seen in section \ref{ssec:eval-age}. In Figure~\ref{fig:irx-beta-grids-3-metallicities} we plot 3 models (using the same parameters as the ones described in section \ref{ssec:eval-age}) which differ only by the metallicity: \textcolor{red}{$Z=0.1$}, \textcolor{green}{$Z=0.02$} and \textcolor{blue}{$Z=0.004$}.

The most metal poor model spans a small UV color range, slightly more than 0.2 dex. The most metal rich model conversely has a much faster evolving ultraviolet color. While this may explain the presence of H$\alpha$ emission in a number of regions it nevertheless still cannot account for the most UV-red regions, assuming a starburst attenuation law.

\begin{figure}[!ht]
 \includegraphics[width=\columnwidth]{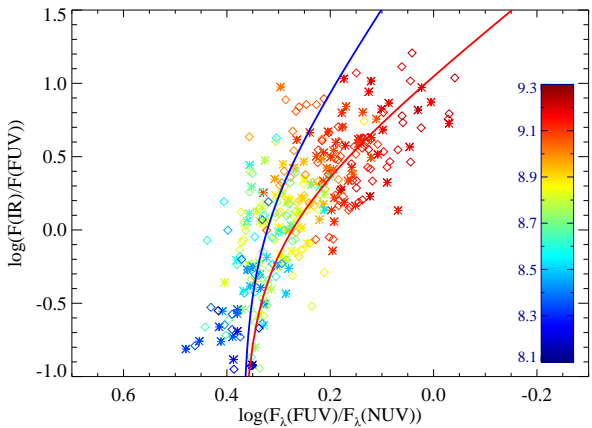}
 \caption{IRX-$\beta$ diagram. The color of the star forming regions indicate their metallicity. The two curves are IRX-$\beta$ relations for the starburst attenuation law (blue) and the foreground SMC (red) extinction law. The age of the instantaneous burst is $6\times10^6$ years old.\label{fig:irx-beta-metals}}
\end{figure}

As we see, the metallicity has an important effect on the location of a star forming region on the IRX-$\beta$ diagram. Another effect of the metallicity is the extinction \citep{calzetti1994a}. Indeed, it is well known that star forming regions with a higher metallicity tend to be more extinguished on average \citep{heckman1998a,boissier2004a,munoz2009a}. An accurate determination of the metallicity of 324 star forming regions in 8 different galaxies would necessitate long spectroscopic observations that are not available. However, we also know that there is a metallicity gradient in spiral disks. While this does not provide the exact metallicity of a star forming region from its deprojected galactocentric distance, it statistically provides a precious information. In order to evaluate the metallicity, we use the metallicity gradients provided by Moustakas et al. (2009, in preparation) and additional data provided in Table~\ref{tab:galaxy-parameters}. The position angles have been calculated from optical images. The corresponding IRX-$\beta$ diagram is presented in Figure~\ref{fig:irx-beta-metals}.

It appears there is indeed a strong correlation between the metallicity and the extinction. The low metallicity regions have a blue ultraviolet color and a low IRX while regions with the highest metallicity have the reddest ultraviolet color and a high IRX. This is consistent with the models presented in Figure~\ref{fig:irx-beta-grids-3-metallicities} where for a given age, a star forming region with a higher metallicity would have a redder ultraviolet color. The Pearson correlation coefficient between the metallicity and IRX is $r=0.80$. The relation is tighter than the one found by \cite{cortese2006a} for optically selected, integrated normal star-forming galaxies, they derived a Spearman correlation coefficient $r_s=0.59$.

\begin{figure}[!ht]
 \includegraphics[width=\columnwidth]{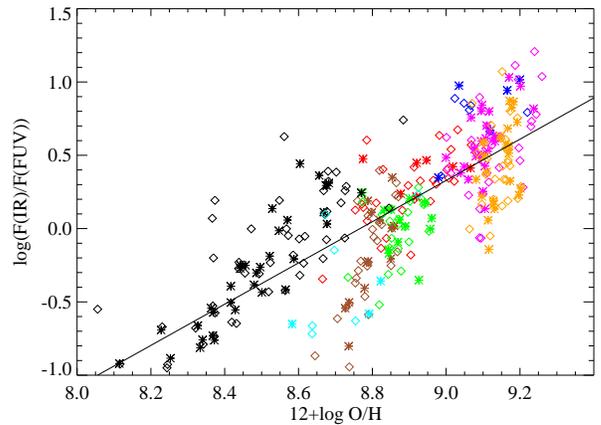}
 \caption{IRX versus the oxygen abundance determined using the galactic metallicity gradient.\label{fig:dist-metal-SB}}
\end{figure}

In order test quantitatively the dependence of the extinction on the metallicity, we plot in Figure~\ref{fig:dist-metal-SB} the IRX value of each star forming region versus the oxygen abundance $12+\log O/H$. 

The Pearson correlation coefficient between the metallicity and IRX is $r=0.80$.

In this section we have examined whether the metallicity can explain the deviation from the IRX-$\beta$ relation. We have shown that the metallicity however is an important parameter to determine the location of the star forming regions in the IRX-$\beta$ diagram. The more metallic they are, the higher IRX and the redder $\beta$ they have, closely following IRX-$\beta$ relations but showing no obvious trend with the perpendicular distance.

\section{Discussion}

Understanding the deviation from the starburst attenuation law in the IRX-$\beta$ diagram is essential in our understanding of star forming galaxies as they provide important insights into the attenuation and reprocessing of starlight. We have shown that the age of the star forming regions -- a debated parameter -- has very little influence on the deviation, which actually necessitates a range of geometries and extinction laws. In addition, the influence of the star formation history and in particular any dilution by an older population should not be neglected. If these parameters are not taken into account properly, the error on the determination star formation rates can be severely impacted.

\subsection{Age of the star forming regions}

We have shown in section \ref{ssec:perp-dist} that there is a weak correlation between the perpendicular distance from the IRX-$\beta$ relation and the age as traced by the H$\alpha$ equivalent width and the $U-B$ color. In all cases, the trends with the age are too weak relative to the scatter in the data to support the notion of aging stellar populations as an important ``second parameter'', at least for the timescales probed by the $U-B$ color and the H$\alpha$ equivalent width. We have seen in Figures~\ref{fig:dist-EW-Ha} and \ref{fig:dist-U-B-corrected} that neither the $U-B$ color nor the H$\alpha$ equivalent width correlate strongly with the perpendicular distance. The absence of correlation is independent of the ``reference'' attenuation curve, be it the starburst or the SMC extcintion law. The reddening of the SED is therefore not due to age effects but is a consequence of another parameter.

It appears that, contrary to what has been found studying integrated galaxies by \cite{kong2004a} for instance -- but in agreement with \cite{johnson2007b} -- there is no obvious trend between the perpendicular distance and the age in the case of individual star forming regions.

\subsection{Necessity of a range of extinction laws\label{ssec:disc-ext-law}}

In this study we have tested 3 different kinds of extinction laws: starburst \citep{calzetti1994a,calzetti2000a,meurer1999a}, foreground SMC \citep{gordon2003a} and mixed SMC. Conventionally the perpendicular distance is derived from the IRX-$\beta$ starburst relation \citep[][for instance]{kong2004a,cortese2006a}. There are several indications that there is a range of extinction laws at work in quiescent star forming galaxies.

As inferred from Figures~\ref{fig:irx-beta}, \ref{fig:UV-color-Av-SB-SMC} and \ref{fig:irx-beta-metals}, both a starburst attenuation curve and an SMC extinction curve with foreground dust are required to span the locus occupied by the data points. Neither curve individually is a perfect match to the data, and both (plus combinations of the two) are required to span a large range of the data locus. Likely this indicates a variation in the applicable dust geometries to the individual regions. 

In section \ref{sec:extinc-laws}, we have shown that assuming a range of models (starburst, foreground SMC assuming $A_V=A_V^{gas}$ or $A_V=0.44\times A_V^{gas}$ -- this ratio is prominently dependent on the dust-stars geometry -- and mixed SMC), it is possible to reproduce simultaneously the UV color and the H$\alpha$ emission of most star-forming galaxies. We assume that the starburst and foreground SMC are the two most ``extreme'' laws at work in quiescent star forming galaxies and that there is a range of dust-stars geometries in between. The two curves (and the SMC needs to be implemented with differential extinction between stellar and nebular emission) bracket most of the range of observed values. A variation of the extinction curve has also been detected for UV-luminous galaxies at $z\sim2$ by \cite{noll2009a}.

\subsection{Influence of the star formation history and dilution by an older population}

Dilution of actively (H$\alpha$ emitting) star forming regions by an older ($\sim300\times10^6$ years) instantaneous burst population is a necessary requirement for a few of the reddest (in UV colors) regions but not for all.

We remove local background from all our star forming regions, thus removing Hubble-time-averaged stellar populations. However, the typical region sampled by our apertures is around 300-500~pc. In this physical region, we can expect multiple populations to be present and possibly contribute to the observed colors. Indeed for the reddest UV color regions, the required mass ratio between an ionizing stellar population and an older, $300\times10^6$~years old stellar population is about $1/5$. This indicates the possibility of 5-6 generations of stars over the past $300\times10^6$~years, within a $500$~pc region. However our data cannot discriminate between this possibility and the alternative of yet more complex star-dust geometries and extinctions than what considered here.

Interestingly we note that very few regions that are detected in the UV do not have any H$\alpha$ counterpart. One of the possible reasons is that regions that are sufficiently old do not emit in H$\alpha$ anymore and tend to be dissolve to become part of the diffuse UV emission from the galaxy. Hence they were not selected and only acted as a contaminant. The fact that we do not find the trend that can be found in integrated galaxies hints that it may be due in reality due to the contamination by older stellar populations. Indeed, as we have seen in this paper the age or the latest generation of stars does not seem to play any significant role. We think the birthrate parameter is not appropriate either in the sense that it is inaccurate as it averages the SFR over the life of the galaxy whereas the typical lifespan of the UV is about $10^9$~years. We therefore think that the fact that few regions are not detected in H$\alpha$ hints in favor of the scenario of the contamination.

Finally, it can be noted that Dale et al. (2009) found a trend between the perpendicular distance and the H$\alpha$ equivalent width calculated using the same method as the one in the paper. The most likely explanation is that the trend is due to the pollution of the underlying continuum by evolved stellar populations. Indeed, their average H$\alpha$ equivalent width are averaged over the galaxy. If we assume that the mean age of star forming regions is the same in all galaxies -- which is is a reasonable assumption for quiescent star forming galaxies --, the the differing amount of underlying stellar populations can create such a correlation: they simultaneously increase the perpendicular distance reddening the UV color and lower the H$\alpha$ equivalent width by increasing the flux of the continuum. In addition, as the R-band emitting and the UV emitting stars have a different life expectancy, it induces some scatter in the relation.

\subsection{Caveats of star formation rate derivations}

From what we have shown previously, the usual star formation rates estimates based on the assumption of the starburst IRX-$\beta$ relation are unreliable in the case of quiescent star forming galaxies. To estimate the star formation rate one approach can be to derive an empirical law from a sample of galaxies, like has been done by \cite{cortese2006a} for instance, however the scatter remains important. 

\begin{figure}[!ht]
 \includegraphics[width=\columnwidth]{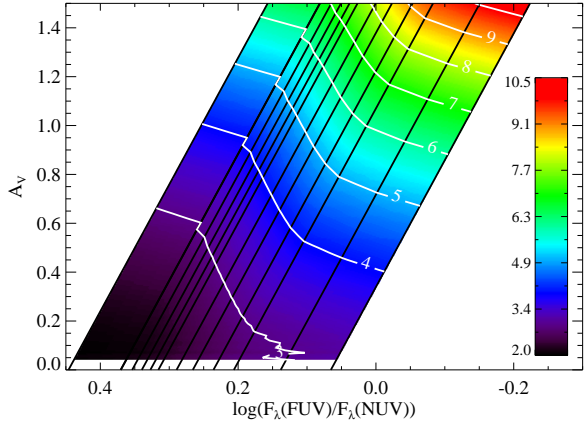}
 \caption{Ratio between the SFR derived from the IRX-$\beta$ relation for the starburst attenuation law (SFR(SB)) and for the foreground SMC extinction law (SFR(SMC)) as a function of the UV color and the extinction (in magnitudes) of the stars. The color indicates the value of the ratio. Assuming an extinction lower than 1.5 magnitude, SFR(SB) can overestimate the SFR by a factor 2 to 10.5 relatively to SFR(SMC). The white lines indicate the value of the ratio. The black lines indicate the track for a given age.\label{fig:ratio-sfr-sb-smc}}
\end{figure}

To determine accurately the actual law, it is crucial to estimate the fraction of the UV color which is contributed by an older population as we have shown in section \ref{ssec:disc-dilution}. Taking into account the age of the stellar population, \cite{cortese2008a} for instance have provided some recipes to correct for the UV extinction, assuming a Large Magellanic Cloud extinction law. Another critical point to derive the actual star formation rate is the choice of an extinction law. Indeed, as we have already mentioned, the effect of the extinction is much more dramatic on the UV color in the case of a SMC law. This can be seen in Figure~\ref{fig:UV-color-Av-SB-SMC} for instance. In Figure~\ref{fig:ratio-sfr-sb-smc} we plot the ratio of the estimated SFR assuming a starburst attenuation law to the estimated SFR assuming a SMC extinction law as a function of the UV color.

We see that indeed the effect on the derived star formation law can be quite dramatic. If we assume a starburst attenuation law, for $A_V=1$ we can overestimate the SFR by a factor up $\sim7$ if the actual extrinction law is foreground SMC like. Conversely, a similar error, but in the sense of underestimating the SFR, will be made if the foreground SMC extinction is adopted, but the actual extinction is described by a starburst attenuation curve.

As we have shown, the age seems to be only a minor factor regarding the location on the IRX-$\beta$ diagram. However, standard star formation rate estimators are significantly age-dependent. Indeed, the \cite{kennicutt1998a} estimators for instance are based on the assumption of a constant star formation rate during $100\times10^6$ years. However if it is sensible for a galaxy-averaged luminosity, it is unlikely that small individual star forming regions undergo such a star formation history. If we assume a foreground SMC, the best-fit age is $6\times10^6$ years.

\subsection{Completeness of the parameter space sampling}

Even if we have selected over 300 star forming regions, the location of each of the region is dependent on the galaxy metallicity for instance. As no such study has been performed before, we can only compare to samples of integrated galaxies. In Figure~\ref{fig:irx-beta} we plot the IRX-$\beta$ diagrams along with data points from SINGS and LVL (Dale et al. 2009).

We see that the IRX range is very similar to that of individual star forming regions in galaxies. Regarding the UV color, it tends to be redder. This is particularly striking for galaxies which have a small IRX lower than $-0.3$, but have red UV colors of $0.24\pm0.07$ compared to $0.37\pm0.05$ for galactic star forming regions. This can be explained by the influence of an older stellar population. Indeed, photometry of integrated galaxies is particularly influenced by a near-ultraviolet emitted older stellar population which reddens the UV color. Conversely, individual star forming regions are relatively less affected by older populations as 1) the specific star formation rate in a star forming region is higher than for an integrated galaxy hence lowering the relative luminosity of any older population 2) the way the background level is measured naturally subtracts the contribution of an older populations in the aperture (see section \ref{ssec:apertures}). Therefore, rather than an undersampling of the parameter space, the fact that we have not detected regions which have simultaneously a red UV color and a low IRX is a consequence of a bias introduced by the emission of older stellar populations in the case of integrated galaxies. As we can see in Figure~\ref{fig:irx-beta}, the envelope of SINGS and LVL galaxies is shifted towards redder colors compared to the individual star forming regions and there is a significant number of galaxies with extremely red UV colors: 9.6\% of the galaxies in the combined LVL and SINGS sample and 1.2\% of the individual star forming regions have a $\log F_\lambda\left(FUV\right)/F_\lambda\left(NUV\right)$ redder than 0.

\section{Summary and conclusions}

We have performed an analysis of {\bf over 300} star forming regions in 8 local quiescent star forming galaxies to garner new insights on the IRX-$\beta$ relation by using Spitzer infrared, GALEX ultraviolet and ground based optical broad band and narrow-band H$\alpha$ data. Specifically we investigate the deviations of star forming regions in normal star forming galaxies from the locus identified by the starburst attenuation curve, in hope of obtaining information on similar deviations by normal star forming galaxies.

The mean age of the stellar population, as traced by the $U-B$ color and by the equivalent width of H$\alpha$, is shown to be relatively uncorrelated or only weakly correlated with the perpendicular distance of the data from the starburst attenuation curve in the IRX-beta diagram. Stellar population ages have been suggested as a potential culprit for the deviations in the case of normal star-forming galaxies \citep[e.g.][via the b-parameter]{kong2004a}, but we do not find evidence for such trend in HII knots of galaxies. Our results are in agreement with similar results, obtained for whole galaxies, by \cite{seibert2005a} and \cite{johnson2007b}, but disagree with recent findings by Dale et al. (2009), also derived for whole galaxies.

We find that a range of dust extinction/attenuation curves and dust/star geometries are actually required to account for the location on the IRX-beta plot of most of our HII knots. Specifically, we find that the starburst attenuation curve and the SMC extinction with foreground geometry bracket the location of most of the data points, not only on the IRX-beta diagram but on other plots reporting colors and other properties of the regions. For the reddest (in UV color) regions, we need to include contributions to the observed colors and fluxes of underlying older ($\sim$ 300 Myr) stellar populations to account for the observed UV and $U-B$ colors. The mass ratio between the young (ionizing) stellar population and the older one is about 1/5.

Individual star forming regions and integrated galaxies do not populate the same locus in the IRX-$\beta$ diagram. Indeed, the former have a bluer UV color and are more tightly correlated. The most likely explanation is that older stellar populations up to $1-2\times10^9$ years old, which have a redder UV color redden the observed UV color of star forming galaxies. To accurately determine the star formation rate, it is crucial to take into account this older population.

Finally, a strong correlation is found with the metallicity, in the sense that the more metal rich regions tend to display redder UV colors and larger IR/UV values, as expected metallicity and dust content correlate with each other.

\acknowledgments

This work has been supported by NASA ADP grant NNX07AN90G.

{\it Facilities:} \facility{Spitzer}, \facility{GALEX}.

\appendix

\section{Apertures}

\begin{figure}[!ht]
 \includegraphics[width=\textwidth]{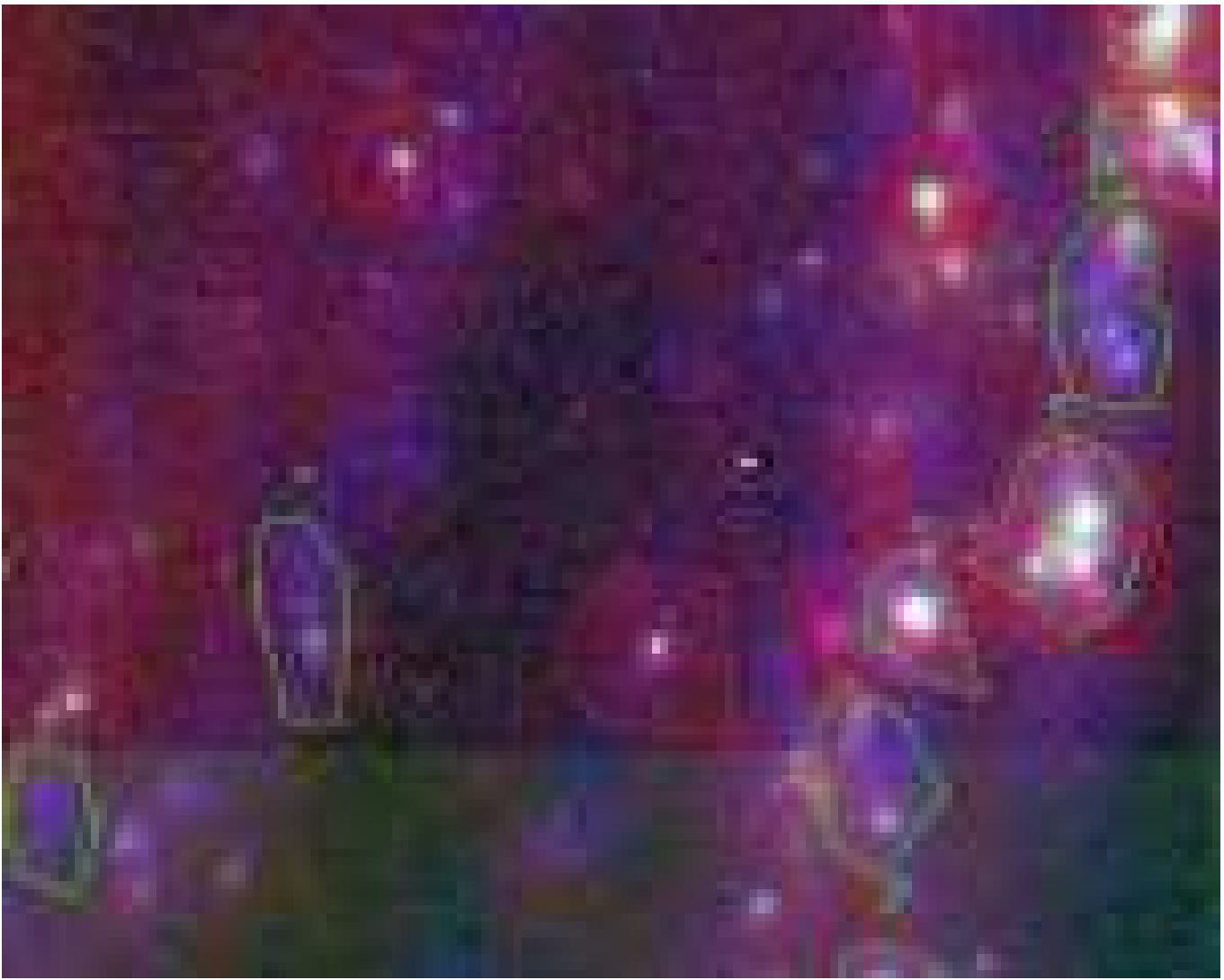}
 \caption{Section of NGC~3031 seen in FUV (blue), H$\alpha$ (green) and 24~$\mu$m (red). The overplotted countours are the apertures made from FUV (green) and 24~$\mu$m (red).\label{fig:apertures}}
\end{figure}

In Figure~\ref{fig:apertures} we see the apertures overplotted on a section of NGC~3031. This illustrates the fact that some regions are selected in FUV only or 24~$\mu$m only. Some regions though are selected from both bands but the shape of the aperture may vary depending on the actual shape of the star-forming region in a given band.

\bibliographystyle{aa}
\bibliography{article}

\end{document}